\documentclass[aip,jmp,10pt]{revtex4-1}
\usepackage{amssymb,amsmath}

\usepackage{mathrsfs}
\usepackage{epsfig}
\usepackage{epstopdf}

\newtheorem{theorem}{Theorem}

\newtheorem{corollary}[theorem]{Corollary}

\newtheorem{definition}[theorem]{Definition}

\newtheorem{lemma}[theorem]{Lemma}

\newtheorem{proposition}[theorem]{Proposition}
\newtheorem{remark}[theorem]{Remark}

\begin{document}

\title{Characteristic Operator Functions for Quantum Input-Plant-Output Models \&
Coherent Control}
\author{John E.~Gough}
\affiliation{Aberystwyth University, Aberystwyth, SY23 3BZ, Wales, United Kingdom}

\begin{abstract}
We introduce the characteristic operator as the generalization of the usual
concept of a transfer function of linear input-plant-output systems to
arbitrary quantum nonlinear Markovian input-output models. This is intended
as a tool in the characterization of quantum feedback control systems that
fits in with the general theory of networks. The definition exploits the
linearity of noise differentials in both the plant Heisenberg equations of
motion and the differential form of the input-output relations.
Mathematically, the characteristic operator is a matrix of dimension equal
to the number of outputs times the number of inputs (which must coincide),
but with entries that are operators of the plant system. In this sense the
characteristic operator retains details of the effective plant dynamical
structure and is an essentially quantum object. We illustrate the relevance
to model reduction and simplification definition by showing that the
convergence of the characteristic operator in adiabatic elimination limit
models requires the same conditions and assumptions appearing in the work on
limit quantum stochastic differential theorems of Bouten and Silberfarb \cite
{Bouten_Silberfarb}. This approach also shows in a natural way that the
limit coefficients of the quantum stochastic differential equations in
adiabatic elimination problems arise algebraically as Schur complements, and
amounts to a model reduction where the fast degrees of freedom are decoupled
from the slow ones, and eliminated.
\end{abstract}

\maketitle

\affiliation{Dept. Physics, Aberystwyth University,
SY23 3BZ, Wales, UK.}

\section{Introduction}

There has been much interest lately in the behavior and control of quantum
linear systems, particularly as these are amenable to transfer matrix
function techniques. In this note, we wish to exploit the structural
features of quantum Markovian models to construct an analogue of the
transfer matrix function for non-linear systems. Coming from the classical
direction there has been fruitful application of operator techniques to
control systems in recent years \cite{FF,BSV,Gohm,Gohm2} employing for
instance characteristic functions techniques, multi-analytic operators and
commutant lifting methods. Here we introduce a natural characteristic
operator function associated with a quantum Markov (or SLH) model.

As in standard quantum mechanics, the model is formulated by representing
physical quantities (observables) as self-adjoint operators on a Hilbert
space. The quantum mechanical system (plant) will have underlying Hilbert
space $\mathfrak{h}$ while the input will be a continuous quantum field with
Hilbert space $\mathfrak{F}$. The coupled model will have joint Hilbert
space $\mathfrak{h}\otimes \mathfrak{F}$, which is also the space on which
the output observables act.

The input-plant-output model can be summarized as 
\begin{eqnarray*}
\mathbf{plant~dynamics} &:&j_{t}\left( X\right) =U\left( t\right) ^{\ast
}\left( X\otimes I\right) U\left( t\right) ; \\
\mathbf{output~process} &:&B_{\mathrm{out},i}\left( t\right) =U\left(
t\right) ^{\ast }\left( I\otimes B_{i}\left( t\right) \right) U\left(
t\right) .
\end{eqnarray*}
where $X$ is an arbitrary plant observable, $B_{i}\left( t\right) $ is a
component of the input field, and $U(t)$ is the unitary entangling the plant
with the portion of the bath that has interacted with it over the time
period $[0,t]$.

\subsection{The ``SLH'' Formalism}

In the following we shall specify to a category of model where $U\left(
\cdot \right) $ is a unitary family of operators on $\mathfrak{h}\otimes %
\mathfrak{F} $, satisfying a differential equation of the form \cite
{HP,partha,Gardiner,Gardiner93} 
\begin{eqnarray}
dU\left( t\right) = \left\{ \sum_{ij}\left( S_{ij}-\delta _{ij}\right)
\otimes d\Lambda _{ij}\left( t\right) +\sum_{i}L_{i}\otimes dB_{i}^{\ast
}\left( t\right) \right. \left. -\sum_{ij}L_{i}^{\ast }S_{ij}\otimes
dB_{j}\left( t\right) +K\otimes dt\right\} U\left( t\right) ,\quad U(0)=I,
\label{HP_QSDE}
\end{eqnarray}
Formally, we can introduce input process $b_{\mathrm{in},i}\left( t\right) $
for $i=1,\cdots ,n$ satisfying singular commutation relations of the form $[
b_{ i}(t),b_{ j}(t^{\prime })^{\ast }]=\delta _{ij}\delta (t-t^{\prime })$,
so that the processes appearing in (\ref{HP_QSDE}) are 
\begin{eqnarray*}
\Lambda _{ij}\left( t\right) \triangleq \int_{0}^{t}b_{i}\left( t^{\prime
}\right) ^{\ast }b_{ j}(t^{\prime })dt^{\prime }, B_{i}\left( t\right)
^{\ast } \triangleq \int_{0}^{t}b_{i}\left( t^{\prime }\right) ^{\ast
}dt^{\prime },\quad B_{j}\left( t\right) \triangleq \int_{0}^{t}b_{
j}(t^{\prime })dt^{\prime }.
\end{eqnarray*}
More exactly, the are rigorously defined as creation and annihilation field
operators on the Boson Fock space $\mathfrak{F}$ over $L_{\mathbb{C}%
^{n}}^{2}\left( \mathbb{R}\right) $. The increments in (\ref{HP_QSDE}) are understood to be
future pointing in the Ito sense. We have the following table of
non-vanishing products 
\begin{eqnarray*}
d\Lambda _{ij}d\Lambda _{kl} &=&\delta _{jk}d\Lambda _{il},\qquad d\Lambda
_{ij}dB_{k}^{\ast }=\delta _{jk}dB_{i}^{\ast } \\
dB_{i}d\Lambda _{kl} &=&\delta _{ik}dB_{l},\qquad dB_{i}dB_{k}^{\ast
}=\delta _{ij}dt.
\end{eqnarray*}

Necessary and sufficient conditions for unitarity \cite{HP,partha} are that
we can collect the coefficients of (\ref{HP_QSDE}) to form a triple $(S,L,H)$%
, which we call the \emph{Hudson-Parthasarathy (HP) parameters}, consisting
of a unitary matrix $S$, a column vector $L$, and a self-adjoint operator $H$%
, 
\begin{eqnarray}
S=\left[ 
\begin{array}{ccc}
S_{11} & \cdots & S_{1n} \\ 
\vdots & \ddots & \vdots \\ 
S_{n1} & \cdots & S_{nn}
\end{array}
\right] , \quad L=\left[ 
\begin{array}{c}
L_{1} \\ 
\vdots \\ 
L_{n}
\end{array}
\right] ,
\end{eqnarray}
with $S_{ij},L_{i},H$ are all operators on $\mathfrak{h}$, and where 
\begin{eqnarray}
K\equiv -\frac{1}{2}\sum_{i}L_{i}^{\ast }L_{i}-iH.
\end{eqnarray}
It has become fashionable to refer to this, plus the related feedback
network models \cite{GoughJamesIEEE09,GJ09}, as the \lq\lq SLH \rq\rq
formalism.

We shall refer to $U(t)$ as the unitary determined by the coupling
parameters $\left( S,L,H\right) $. In differential form, the
input-plant-output model then becomes \cite{HP,partha}

\bigskip

\noindent \textbf{plant dynamical (Heisenberg) equation:} 
\begin{eqnarray}
dj_{t}\left( X\right) =j_{t}\left( \mathscr{L}X\right)
dt+\sum_{i}j_{t}\left( \mathscr{M}_{i}X\right) dB_{i}^{\ast }\left( t\right)
+\sum_{i}j_{t}\left( \mathscr{N}_{i}X\right) dB_{i}\left( t\right)
+\sum_{j,k}j_{t}\left( \mathscr{S}_{jk}X\right) d\Lambda _{jk}\left(
t\right) ;
\end{eqnarray}
\bigskip

Here 
\begin{eqnarray}
\mathscr{L}X &=&\frac{1}{2}\sum_{i}L_{i}^{\ast }\left[ X,L_{i}\right] +\frac{%
1}{2}\sum_{i}\left[ L_{i}^{\ast },X\right] L_{i}-i\left[ X,H\right] , \quad 
\text{(the Lindbladian!)}, \\
\mathscr{M}_i X &=& S_{ji}^\ast [X,L_j ], \\
\mathscr{N}_i X &= & [L_k^\ast , X ] S_{ki}, \\
\mathscr{S}_{ik} X &=& S_{ji}^\ast XS_{jk} - \delta_{ik} X .
\end{eqnarray}

\noindent \textbf{input-output relations:}

\begin{eqnarray}
dB_{\mathrm{out},i}\left( t\right) =j_{t}\left( S_{ik}\right) dB_{k}\left(
t\right) +j_{t}\left( L_{i}\right) dt.
\end{eqnarray}

\subsection{Linear Quantum Models}

\label{sec:LQM} If we specify to a system of quantum mechanical oscillators
with modes $a_{1},\cdots ,a_{m}$ satisfying canonical commutation relations 
\begin{eqnarray}
\left[ a_{\alpha },a_{\beta }^{\ast }\right] =\delta _{\alpha \beta }
\end{eqnarray}
then we obtain a linear dynamical model with the prescription 
\begin{eqnarray}
S_{ij}=D_{ij},\,L_{i}=\sum_{\alpha =1}^{m}C_{i\alpha }a_{\alpha
},\,H=\sum_{\alpha ,\beta }a_{\alpha }^{\ast }\omega _{\alpha \beta
}a_{\beta }.
\end{eqnarray}
Specifically, the plant dynamics and input-output relations are affine
linear in the mode variables $a_{i}$: 
\begin{eqnarray*}
da_{\alpha }\left( t\right) &=&\sum_{\beta }A_{\alpha \beta }a_{\beta
}\left( t\right) dt+\sum_{i}B_{\alpha i}dB_{i}\left( t\right) ; \\
dB_{\mathrm{out},i}\left( t\right) &=&\sum_{\beta }C_{i\beta }a_{\beta
}\left( t\right) dt+\sum_{k}D_{ik}dB_{k}\left( t\right) .
\end{eqnarray*}
where, setting $D=\left[ D_{ij}\right] \in \mathbb{C}^{n\times n}$, $C=\left[
C_{i\alpha }\right] \in \mathbb{C}^{n\times m}$ and $\Omega =\left[ \omega
_{\alpha \beta }\right] \in \mathbb{C}^{m\times m}$, we have 
\begin{eqnarray}
A=-\frac{1}{2}C^{\ast }C-i\Omega ,\quad B=-C^{\ast }S.
\end{eqnarray}

In turn, a model having this specific structure is said to be physically
realizable. The transfer matrix associated with the linear dynamics is then
defined to be \cite{YK1,YK2,GGY08} 
\begin{eqnarray}
T(s) = \left[ 
\begin{tabular}{l|l}
$A$ & $B$ \\ \hline
$C$ & $D$%
\end{tabular}
\right] (s)\triangleq D+C\left( sI-A\right) ^{-1}B,  \label{eq:T}
\end{eqnarray}
and there exists a well-established literature developing control theory
from analysis of these functions.

The definition here leads to transfer functions that are positive real
functions of the complex variable $s$, and they model passive systems. The
generalization to active linear models, which we do not need here, is given
in \cite{GJN10}.

\subsection{Characteristic Operators}

In the mathematical formulation of open quantum Markov systems, a natural
role is played by the model matrix, introduced in \cite{GJ09}, 
\begin{equation}
\mathbf{V} =\left[ 
\begin{tabular}{ll}
$-\frac{1}{2}L^{\ast }L-iH$ & $-L^{\ast }S$ \\ 
$L$ & $S$%
\end{tabular}
\right] .  \label{model_matrix}
\end{equation}
We now use it as the basis for the definition of an operator-valued
generalization of the characteristic function.

\begin{definition}[The Characteristic Operator]
For given $\left( S,L,H\right) $ we define the corresponding characteristic
operator by 
\begin{equation}
\mathscr{T}(s)\triangleq \left[ 
\begin{tabular}{l||l}
$-\frac{1}{2}L^{\ast }L-iH$ & $-L^{\ast }S$ \\ \hline\hline
$L$ & $S$%
\end{tabular}
\right] (s)=S-L(sI+\frac{1}{2}L^{\ast }L+iH)^{-1}L^{\ast }S.
\label{characteristic op}
\end{equation}
We shall often write $\mathscr{T}_{(S,L,H)}$ for emphasis.
\end{definition}

\begin{lemma}
The characteristic operator $\mathscr{T}(s)$ is a bounded operator for Re$%
\,s>0$. For all $\omega \in \mathbb{R}$, such that $i\omega$ lies in the
resolvent set of $K=-\frac{1}{2}L^{\ast }L-iH$ (that is, whenever $i\omega
-K $ is invertible), we have $\mathscr{T}($i$\omega )$ well-defined and
unitary: 
\begin{eqnarray}
\mathscr{T}(i\omega )^{\ast }\mathscr{T}(i\omega )=\mathscr{T}(i\omega )%
\mathscr{T}(i\omega )^{\ast }=I.
\end{eqnarray}
\end{lemma}

The proof follows mutatis mutandis of the proof of an analogous result in 
\cite{GGY08}.

\subsection{Examples}

\subsubsection{Lossless System}

Suppose that we have no coupling $L=0$ then the characteristic operator is $%
\mathscr{T} (s) \equiv S$, constant. This is true even if $H$ is non-zero.
Without coupling, we cannot infer anything about the system Hamiltonian.

\subsubsection{Quantum Linear Passive System}

For the model considered in subsection \ref{sec:LQM} we have 
\begin{eqnarray}
\mathscr{T} (s) = S - Ca \frac{1}{s-a^\ast A a }a^\ast C^\ast S.
\label{eq:QLPS}
\end{eqnarray}
where $a^\ast = [a_1^\ast , \cdots , a_m^\ast ]$. For the $m=n=1$ case we
have explicitly 
\begin{eqnarray}
\mathscr{T} (s) = S - C \frac{1}{s- A (N+1) } C^\ast S,
\end{eqnarray}
where $N=a^\ast a$ is the number operator for the single mode. In fact, we
see that 
\begin{eqnarray}
\langle \mathscr{T} (s) \rangle_{\mathrm{vac}}= T(s),
\end{eqnarray}
where $T(s)$ is the transfer function (\ref{eq:T}). The same vacuum
expectation is obtained for the cases $n,m $ greater than one.

\subsubsection{Qubit Example}

A simple example is a qubit system with master equation 
\begin{eqnarray}
\frac{d}{dt}\varrho =\mathscr{D}_{L}\varrho +i\left[ \varrho ,H\right]
\end{eqnarray}
where $\mathscr{D}_{L}\varrho =L\varrho L^{\ast }-\frac{1}{2}\left\{ \varrho
L^{\ast }L+L^{\ast }L\varrho \right\} $ and we set $L=\sqrt{\gamma \left(
n+1\right) }\sigma _{-}+\sqrt{\gamma n}\sigma _{+}$, $H=\omega \sigma _{z}$.
This models a qubit in a thermal bath with $0\leq n\leq 1$ being the
equilibrium occupancy of the state $|\uparrow \rangle $ in the presence of
the oscillation $\omega \sigma _{z}$. We shall take the scattering to be by
a polarization-dependent phase 
\begin{eqnarray}
S=e^{i\varphi _{+}}|\uparrow \rangle \langle \uparrow |+e^{i\varphi
_{-}}|\downarrow \rangle \langle \downarrow |.
\end{eqnarray}
In the $\sigma _{z}$-basis $|\uparrow \rangle =\left[ 
\begin{array}{c}
1 \\ 
0
\end{array}
\right] ,|\downarrow \rangle =\left[ 
\begin{array}{c}
0 \\ 
1
\end{array}
\right] ,\,K=\left[ 
\begin{array}{cc}
-\frac{1}{2} \gamma \left( n+1\right) -i\omega & 0 \\ 
0 & -\frac{1}{2} \gamma n + i\omega
\end{array}
\right] $ and the characteristic operator explicitly is

\begin{eqnarray}
\mathscr{T}\left( s\right) =\left[ 
\begin{array}{cc}
\frac{s-\frac{1}{2}\gamma n-i\omega }{s+\frac{1}{2}\gamma n+i\omega }%
e^{i\varphi _{+}} & 0 \\ 
0 & \frac{s-\frac{1}{2}\gamma (n+1)+i\omega }{s+\frac{1}{2}\gamma
(n+1)-i\omega }e^{i\varphi _{-}}
\end{array}
\right] .
\end{eqnarray}
The characteristic operator is diagonal in the basis $\left\{ |\uparrow
\rangle ,|\downarrow \rangle \right\} $, but this would no longer be true if 
$\left[ S,\sigma _{z}\right] \neq 0$.

\subsubsection{Opto-Mechanical Example}

\begin{figure}[htbp]
\centering
\includegraphics[width=0.40\textwidth]{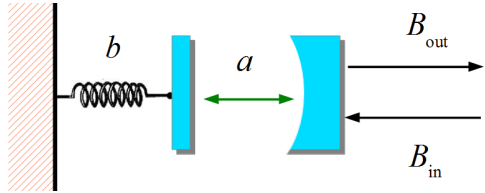}
\caption{(color online) A mechanical mode (moveable mirror) coupled to an
open cavity.}
\label{fig:optomech}
\end{figure}

We consider a model of a cavity mode $a$ between a fixed leaky mirror and a
perfect mirror with quantum mechanical position $X=b+b^{\ast }$, see Fig. 
\ref{fig:optomech}. The SLH model takes the form 
\begin{equation}
S=1,\quad L=\sqrt{\gamma }a,\quad H=\Delta a^{\ast }a+\omega _{0}b^{\ast
}b+gXa^{\ast }a,  \label{eq:SLH_Milburn}
\end{equation}
where $\gamma $ is the damping to the input field at the leaky mirror, $%
\Delta $ is the cavity detuning, $\omega _{0}$ is the harmonic frequency of
the mirror, and $g$ is the coupling strength associated with mirror-mode
interaction. Note that the interaction $g_{0}Xa^{\ast }a$ couples the
position of the mirror to the cavity mode photon number in accordance with
the notion of radiation pressure. This is a standard opto-mechanical model
and we obtain the Langevin equations 
\begin{eqnarray*}
dj_{t}(a) &=&-\left( \frac{1}{2}\gamma +i\Delta +ig_{0}X\right) j_{t}(a)\,dt-%
\sqrt{\gamma }dB\left( t\right) , \\
dj_{t}\left( b\right) &=&-i\omega _{0}j_{t}(b)\,dt-g_{0}j_{t}\left( a^{\ast
}a\right) \,dt.
\end{eqnarray*}

A simplifying assumption is that the mechanical processes are much slower
than the optical ones, in which case we set $\omega _{0}\equiv 0$. The
characteristic operator in this case is 
\[
\mathscr{T}_{\text{optomech}}\left( s\right) =\mathscr{T}_{\left( I,\sqrt{%
\gamma }a,\left( \Delta +gX\right) a^{\ast }a\right) }(s) 
\]
This is recognizable as the characteristic operator of a quantum linear
passive system as in (\ref{eq:QLPS}), but with the operator $A$ taking the
form $A=-\left( \frac{1}{2}\gamma +i\Delta +igX\right) $. That is, $A$ is no
longer scalar valued, but depends explicitly on the position observable $X$
of the mirror. Note that $A$ is still strictly Hurwitz since $X$ is
self-adjoint. We remark that position dependent transfer functions have been
proposed for single photon input-output models for this type of model with
one-particle fields related by \cite{Akrametal} 
\[
\xi _{\text{out}}\left( s\right) =\frac{s-\frac{1}{2}\gamma -i(\Delta +gX)}{%
s+\frac{1}{2}\gamma -i(\Delta +gX)}\xi _{\text{in}}\left( s\right) , 
\]
and here the transfer function corresponds to the partial trace of $%
\mathscr{T}_{\text{optomech}}\left( s\right) $ over the vacuum state of the
cavity.

\subsection{Properties of the Characteristic Operator}

\begin{lemma}[All-Pass Representation]
The characteristic operator admits the following ``all-pass''
representation: 
\begin{eqnarray}
\mathscr{T}(s)=\frac{1-\frac{1}{2}\Sigma \left( s\right) }{1+\frac{1}{2}%
\Sigma \left( s\right) }S,
\end{eqnarray}
where $\Sigma \left( s\right) =L\left( s+iH\right) ^{-1}L^{\ast }$.
\end{lemma}

This is proved in \cite{GZ_mode}, and we recall briefly the proof.

\noindent \textbf{Proof} An application of the Woodbury matrix identity \cite
{Woodbury} $\left( A+UCV\right) ^{-1}=A^{-1}-A^{-1}U\left(
C^{-1}+VA^{-1}U\right) ^{-1}VA^{-1}$ shows that 
\begin{eqnarray}
\frac{1}{sI+iH+\frac{1}{2}L^{\ast }L}=\frac{1}{sI+iH}-\frac{1}{2}\frac{1}{%
sI+iH}\frac{1}{1+\frac{1}{2}L(sI+iH)^{-1}L^{\ast }}\frac{1}{sI+iH}.
\end{eqnarray}
Substituting into $\left( \ref{characteristic op}\right) $ then gives the
above relation after some straightforward algebra. $\square $

Note that $\Sigma \left( i\omega \right) ^{\ast }=-\Sigma \left( i\omega
\right) $ for real $\omega $, so that we could alternatively have deduced
unitarity by a Cayley transformation argument.

\begin{corollary}
Suppose that the model parameters satisfy the condition $[ L,H] \equiv 0 $,
then the characteristic operator takes the form 
\begin{eqnarray}
\mathscr{T}(s)=\frac{s-\frac{1}{2}LL^{\ast }+iH}{s+\frac{1}{2}LL^{\ast }+iH}%
S.
\end{eqnarray}
\end{corollary}

The condition $[ L,H] \equiv 0 $ arises as the QND condition for measurement
disturbance in the sense of Braginsky \cite{Brag}.

\begin{remark}[Equivalence to passive systems]
\label{ref:passive_equiv} For a finite-dimensional system, say with Hilbert
space $\mathfrak h = \mathbb{C}^m$, we may fix an orthonormal basis of $m$
vectors for $\mathfrak h$. In this representation, we may describe $H$ as an 
$m\times m$ matrix which we denote as $\Omega \in \mathbb{C}^{m \times m}$. The
coupling operator $L$ is then a column vector of $n$ operators, each
represented as an $m \times m$ matrix, so that $L$ may be represented as an $%
nm \times m$ matrix which we denotes as $C\in \mathbb{C}^{nm \times m}$. In
this manner, $S$ becomes a complex valued matrix $D \in \mathbb{C}^{nm \times
nm}$. We then have the equivalence

\begin{equation}
\mathscr{T}(s) = \left[ 
\begin{tabular}{l||l}
$-\frac{1}{2}L^{\ast }L-iH$ & $-L^{\ast }S$ \\ \hline\hline
$L$ & $S$%
\end{tabular}
\right] (s) \equiv \left[ 
\begin{tabular}{l|l}
$A$ & $B$ \\ \hline
$C$ & $D$%
\end{tabular}
\right] (s) ,  \label{eq:equiv}
\end{equation}
where $A= -\frac{1}{2} C^\ast C -i \Omega$ and $B= -C^\ast D$. In this was
we realise the characteristic function as the transfer operator of a linear
passive system $A,B,C,D$, structurally similar to those considered in
subsection \ref{sec:LQM}, with a state space of $m$ dimensions and $nm$
inputs.
\end{remark}

\subsection{Stratonovich Form of the Characteristic Operator}

We show now that the characteristic operator function can be described in
terms of the coefficient operators in the Stratonovich QSDE.

The Stratonovich differential is defined using the midpoint rule convention
which leads to the algebraic rule \cite{Gough7,GoughWong-Zakai} 
\[
dX_{t}\circ Y_{t}\triangleq dX_{t}\,Y_{t}+\frac{1}{2}dX_{t}\,dY_{t} 
\]
It can then be shown that the Stratonovich form of the QSDE (\ref{HP_QSDE})
takes the form 
\[
dU\left( t\right) =\left\{ \sum_{ij}E_{ij}\otimes d\Lambda _{ij}\left(
t\right) +\sum_{i}E_{i0}\otimes dB_{i}^{\ast }\left( t\right)
+\sum_{j}E_{0j}\otimes dB_{j}\left( t\right) +E_{00}\otimes dt\right\} \circ
U\left( t\right) ,\quad U(0)=I, 
\]
with $E_{ij}^{\ast }=E_{ji}$, $E_{i0}^{\ast }=E_{0i}$ and $E_{00}^{\ast
}=E_{00}$. It is convenient to collect all the coefficients into a
(Hermitean) matrix 
\begin{equation}
\mathbf{E}=\left[ 
\begin{array}{cc}
E_{00} & E_{0\ell } \\ 
E_{\ell 0} & E_{\ell \ell }
\end{array}
\right]  \label{eq: model E}
\end{equation}
The components of $\mathbf{E}$ are related to the $(S,L,H)$ by the
transformation \cite{Gough7,GoughWong-Zakai}: $S=[S_{ij}]_{1\leq i,j\leq n}$
is the Cayley transform $E_{\ell \ell }=[E_{ij}]_{1\leq i,j\leq n}$, 
\begin{equation}
S=\frac{1-\frac{i}{2}E_{\ell \ell }}{1+\frac{i}{2}E_{\ell \ell }},  \label{S}
\end{equation}
and therefore $S$ is unitary, while 
\begin{equation}
L=i\frac{1}{1+\frac{i}{2}E_{\ell \ell }}E_{\ell 0},\quad H=E_{00}+\frac{1}{2}%
\mathrm{Im}\left\{ E_{0\ell }\frac{1}{1+\frac{i}{2}E_{\ell \ell }}E_{\ell
0}\right\}  \label{L's}
\end{equation}
with $H$ self-adjoint. Note that the operator $K$ is then given by 
\[
K\equiv -iE_{00}-\frac{1}{2}E_{0\ell }\frac{1}{1+\frac{i}{2}E_{\ell \ell }}%
E_{\ell 0}. 
\]

\begin{lemma}[Stratonovich form of the Characteristic Operator]
We may write the characteristic operator in terms of the coefficients making
up the Stratonovich matrix $\mathbf{E}$ (\ref{eq: model E}) as 
\begin{eqnarray}
\mathscr{T}\left( s\right) \equiv \frac{I-\frac{i}{2}E_{\ell \ell }-\frac{1}{%
2}E_{\ell 0}\frac{1}{s+iE_{00}}E_{0\ell }}{I+\frac{i}{2}E_{\ell \ell }+\frac{%
1}{2}E_{\ell 0}\frac{1}{s+iE_{00}}E_{0\ell }}.  \label{eq:T_Strat}
\end{eqnarray}
\end{lemma}

\noindent \textbf{Proof} We have explicitly that 
\[
\mathscr{T}\left( s\right) =\frac{I-\frac{i}{2}E_{\ell \ell }}{I+\frac{i}{2}%
E_{\ell \ell }}-\frac{1}{I+\frac{i}{2}E_{\ell \ell }}E_{\ell 0}\frac{1}{%
s+iE_{00}+\frac{1}{2}E_{0\ell }\frac{1}{1+\frac{i}{2}E_{\ell \ell }}E_{\ell
0}}E_{0\ell }\frac{1}{I+\frac{i}{2}E_{\ell \ell }} 
\]
The Woodbury matrix identity \cite{Woodbury} with $A=I+\frac{i}{2}E_{\ell
\ell },U=\frac{1}{\sqrt{2}}E_{\ell 0},V=\frac{1}{\sqrt{2}}E_{0\ell
},C=\left( s+iE_{00}\right) ^{-1}$ shows that 
\[
\frac{1}{I+\frac{i}{2}E_{\ell \ell }+\frac{1}{2}E_{\ell 0}\frac{1}{s+iE_{00}}%
E_{0\ell }}=\frac{1}{2}\left( \mathscr{T}\left( s\right) +I\right) . 
\]
Rearranging for then gives the desired result. $\square $

Note that we have the correct limit $\lim_{\left| s\right| \rightarrow
\infty }=\frac{1-\frac{i}{2}E_{\ell \ell }}{1+\frac{i}{2}E_{\ell \ell }}=S$.

Suppose that we have $E_{00}=kF_{00}$, $E_{\ell 0}=kF_{\ell 0}$ and $E_{\ell
\ell }=F_{\ell \ell }$ independent of $k$, then the associated transfer
operator $\mathscr{T}_{k}\left( s\right) $ has the well-defined limit 
\[
\lim_{k\rightarrow \infty }\mathscr{T}_{k}\left( s\right) =\widehat{S}, 
\]
provided that $F_{00}$ is invertible. Here $\widehat{S}=\frac{1-\frac{i}{2}\widehat{E%
}_{\ell \ell }}{1+\frac{i}{2}\widehat{E}_{\ell \ell }}$ with $\widehat{E}_{\ell \ell
}=F_{\ell \ell }-F_{\ell 0}\left( F_{00}\right) ^{-1}F_{0\ell }$. This
limit, which corresponds physically to high-energy and strong damping, leads
to a purely scattering model but with a shifted scattering matrix $\widehat{S}$.
We shall study more general examples of this type of scaling leading to SLH
models with nontrivial couplings $\widehat{L}$ and and Hamiltonians $\widehat{H}$.

\section{Model Simplification and Reduction}

As we have seen, the characteristic operator for a system with underlying
Hilbert space $\mathfrak{h}$ with $n$ inputs is a function taking values in $%
\mathfrak{B}\left( \mathfrak{h}\right) \otimes \mathbb{C}^{n \times n} $, the
set of $n\times n$ matrices with entries in $\mathfrak{B}\left( \mathfrak{h}%
\right)$, the bounded operators on $\mathfrak{h}$.

Let $A$ and $B$ be models with the same input dimension $n$ and having
coefficient parameters $\left( S_{A},L_{A},H_{A}\right) $ and $\left(
S_{B},L_{B},H_{B}\right) $ respectively. We may \emph{cascade} the systems
by feeding the output of $A$ and input to $B$ and in the instantaneous
feedforward limit we get the model $B\vartriangleleft A$ on $\mathfrak{h}=%
\mathfrak{h}_{B}\otimes \mathfrak{h}_{A}$ with parameters given by the
series product, see \cite{GoughJamesIEEE09} and \cite{GJ09}, $\left(
S_{B}\otimes S_{A},L_{B}\otimes I_{A}+S_{B}\otimes L_{A},H_{B}\otimes
I_{A}+I_{B}\otimes H_{A}+\mathrm{Im}\left\{ L_{B}^{\ast }S_{B}\otimes
L_{A}\right\} \right) $. In this case we typically have 
\begin{eqnarray}
\mathscr{T}_{B\vartriangleleft A}\left( z\right) \neq \mathscr{T}\left(
z\right) _{B}\otimes \mathscr{T}_{A}\left( z\right) .
\end{eqnarray}
(Here we employ the shorthand $S_{B}\otimes S_{A}$ for the matrix with $j,k$%
-entries $\sum_{l=1}^{n}\left[ S_{B}\right] _{jl}\otimes \left[ S_{A}\right]
_{lk}$, etc.)

Thus characteristic function for cascaded systems is not naturally the
product of their characteristic operators. For cascaded classical systems,
the state spaces take the form $\mathcal{X}_A$ and $\mathcal{X}_B$ so that
the combined state space is the direct sum $\mathcal{X}_B \oplus \mathcal{X}%
_A $. The rule in quantum theory is that the combined Hilbert state space
for the cascaded systems is the tensor product and not the direct sum. (Note
that for quantum linear systems, the Hilbert space is the Fock space $%
\mathfrak{h} = \Gamma (\mathcal{X})$ over $\mathcal{X}$, and for combined
linear systems we have $\Gamma (\mathcal{X}_A ) \otimes \Gamma (\mathcal{X}%
_B) \cong \Gamma (\mathcal{X}_A \oplus \mathcal{X}_B )$, which is the usual
rule for Fock spaces \cite{partha}. In this way the usually cascade rule
re-emerges for the corresponding transfer functions \cite{GGY08}.)

With this observation, we see that model reduction techniques based around
the characteristic operator should involve direct sum decompositions, say 
\begin{eqnarray}
\mathfrak{h}=\mathfrak{h}_{\mathsf{1}}\oplus \mathfrak{h}_{\mathsf{2}}
\end{eqnarray}
into orthogonal subspaces. Each of the coefficients $X=S_{jk},L_{j},H$,
etc., can be represented as 
\begin{eqnarray}
X=\left[ 
\begin{array}{cc}
X_{\mathsf{11}} & X_{\mathsf{12}} \\ 
X_{\mathsf{21}} & X_{\mathsf{22}}
\end{array}
\right]
\end{eqnarray}
where $X_{\mathsf{ab}}$ maps from $\mathfrak{h}_{\mathsf{b}}$ to $%
\mathfrak{h}_{\mathsf{a}}$. The characteristic operator may similarly be
decomposed as

\begin{eqnarray}
\mathscr{T} ( s ) =\left[ 
\begin{array}{cc}
\mathscr{T}_{\mathsf{11}}(s) & \mathscr{T}_{\mathsf{12}}(s) \\ 
\mathscr{T}_{\mathsf{21}}(s) & \mathscr{T}_{\mathsf{22}}(s)
\end{array}
\right]
\end{eqnarray}
with 
\begin{eqnarray}
\mathscr{T}_{\mathsf{ab}}(s)\equiv \left\{ \delta _{\mathsf{ac}}-L_{\mathsf{%
ad}}\left[ \frac{1}{s-K}\right] _{\mathsf{de}}L_{\mathsf{ce}}^{\ast
}\right\} S_{\mathsf{cb}}.
\end{eqnarray}
(Here we have the convention that repeated \textsf{sans serif} indices are
summed over the range \textsf{1} and \textsf{2}. We also adopt the notation
that $S_{jk}$ is the $\mathfrak{B}\left( \mathfrak{h}\right) $-valued output 
$j$, input $k$ entry of $S$, while $S_{\mathsf{ab}}$ is the component of $S$
mapping from $\mathfrak{h}_{\mathsf{b}}$ to $\mathfrak{h}_{\mathsf{a}}$,
which is an $n\times n$ matrix of maps from $\mathfrak{h}_{\mathsf{b}}$ to $%
\mathfrak{h}_{\mathsf{a}}$. Similarly $L_{\mathsf{ad}}$ is the $n$-column
vector of maps from $\mathfrak{h}_{\mathsf{b}}$ to $\mathfrak{h}_{\mathsf{a}%
} $.)

Using the Schur-Feshbach identity we may write the resolvent $\frac{1}{s-K}$
as

\begin{eqnarray}
\left[ 
\begin{array}{cc}
s-K_{\mathsf{11}} & -K_{\mathsf{12}} \\ 
-K_{\mathsf{21}} & s-K_{\mathsf{22}}
\end{array}
\right] ^{-1}=\left[ 
\begin{array}{ll}
\widehat{\Delta}_{\mathsf{11}}\left( s\right) & \widehat{\Delta}_{\mathsf{12}}\left(
s\right) \\ 
\widehat{\Delta}_{\mathsf{21}}\left( s\right) & \widehat{\Delta}_{\mathsf{22}}\left(
s\right)
\end{array}
\right]
\end{eqnarray}
where, introducing 
\begin{equation}
\widehat{K}_{\mathsf{11}}\left( s\right) =K_{\mathsf{11}}+K_{\mathsf{12}}\frac{1%
}{s-K_{\mathsf{22}}}K_{\mathsf{21}}  \label{K^hat (s)}
\end{equation}
and $\Delta _{\mathsf{22}}\left( s\right) =\frac{1}{s-K_{\mathsf{22}}}$, we
have 
\begin{eqnarray*}
\widehat{\Delta}_{\mathsf{11}}\left( s\right) &=&\frac{1}{s-\widehat{K}_{\mathsf{11}%
}\left( s\right) } \\
\widehat{\Delta}_{\mathsf{12}}\left( s\right) &=&\widehat{\Delta}_{\mathsf{11}%
}\left( s\right) K_{\mathsf{12}}\Delta _{\mathsf{22}}\left( s\right) \\
\widehat{\Delta}_{\mathsf{21}}\left( s\right) &=&\Delta _{\mathsf{22}}\left(
s\right) K_{\mathsf{21}}\widehat{\Delta}_{\mathsf{11}}\left( s\right) \\
\widehat{\Delta}_{\mathsf{22}}\left( s\right) &=&\Delta _{\mathsf{22}}(s)+\Delta
_{\mathsf{22}}\left( s\right) K_{\mathsf{21}}\widehat{\Delta}_{\mathsf{11}%
}\left( s\right) K_{\mathsf{12}}\Delta _{\mathsf{22}}\left( s\right) .
\end{eqnarray*}
The blocks of the characteristic operator partitioned with respect to the
direct sum $\mathfrak{h}=\mathfrak{h}_{\mathsf{1}}\oplus \mathfrak{h}_{%
\mathsf{2}}$\ are then 
\begin{eqnarray}
\mathscr{T}_{\mathsf{ab}}(s)\equiv \left\{ \delta _{\mathsf{ac}}-L_{\mathsf{%
ad}}\widehat{\Delta}_{\mathsf{de}}(s)L_{\mathsf{ce}}^{\ast }\right\} S_{\mathsf{%
cb}}.
\end{eqnarray}

\begin{definition}
Given the direct sum $\mathfrak{h}=\mathfrak{h}_{\mathsf{1}}\oplus %
\mathfrak{h}_{\mathsf{2}}$, we say that orthogonal subspaces $\mathfrak{h}_{%
\mathsf{1}}$ and $\mathfrak{h}_{\mathsf{2}}$ are decoupled if the
characteristic operator takes the block diagonal form 
\begin{eqnarray}
\mathscr{T} (s) =\left[ 
\begin{array}{cc}
\mathscr{T}_{\mathsf{11}}(s) & 0 \\ 
0 & \mathscr{T}_{\mathsf{22}}(s)
\end{array}
\right] ,
\end{eqnarray}
that is $\mathscr{T}_{\mathsf{21}}(s)=0$ and $\mathscr{T}_{\mathsf{12}}(s)=0$%
.
\end{definition}

\bigskip

We note that if $V$ is a unitary on the system space, then the basic unitary
rotation behaviour for characteristic operators is 
\begin{eqnarray}
\mathscr{T}_{(V^{\ast }SV,V^{\ast }LV,V^{\ast }HV)} \equiv V^{\ast }%
\mathscr{T}_{(S,L,H)}V
\end{eqnarray}

However we note following result, which is easily derived.

\begin{proposition}
For any unitary $V$ on the plant Hilbert space, the HP parameters $\left(
S,LV,V^{\ast }HV\right) $ generate the same characteristic operator as $%
\left( S,L,H\right) $. More generally we have the following invariance
property of the characteristic function: 
\begin{eqnarray}
\left[ 
\begin{tabular}{l||l}
$A$ & $B$ \\ \hline\hline
$C$ & $D$%
\end{tabular}
\right] =\left[ 
\begin{tabular}{l||l}
$V^{\ast }AV$ & $V^{\ast }B$ \\ \hline\hline
$CV$ & $D$%
\end{tabular}
\right] .
\end{eqnarray}
\end{proposition}

Therefore, while the characteristic operator is a quantum object - for $n$
inputs, it is an $n\times n$ matrix with entries that are operators on the
plant space - its dependence on the plant operators is only up to a unitary
equivalence as outlined in the proposition.

\begin{definition}
Let $\left( S,L,H\right) $ be given HP parameters for a fixed plant Hilbert
space $\mathfrak{h}$. If $\left( S^{\prime },L^{\prime },H^{\prime }\right) $
are HP parameters for a proper subspace $\mathfrak{h}^{\prime }$ of the
plant space then we say that $\left( S^{\prime },L^{\prime },H^{\prime
}\right) $ is a reduced model of $\left( S,L,H\right) $ if we have 
\begin{eqnarray}
\mathscr{T}_{\left( S,L,H\right) }= \left[ 
\begin{array}{cc}
\mathscr{T}_{\left( S^{\prime },L^{\prime},H^{\prime }\right) } & 0 \\ 
0 & I
\end{array}
\right] ,
\end{eqnarray}
with respect to the decomposition $\mathfrak{h} = \mathfrak{h}^\prime \oplus
(\mathfrak{h}^\prime)^\perp$. A reduced model is minimal if it allows no
further model reduction.
\end{definition}

\subsection{Examples}

\subsubsection{Detuned Two-Level Atom}

\label{subsubsec:detuned} As a simple toy model, let us consider a two-level
atom with ground and excited states states $| g \rangle$ and $|e \rangle$.
We fix the open system as being a single input model with $S=I$, $L=\sqrt{%
\gamma} \sigma_z + \sqrt{\kappa} \sigma_-$ and Hamiltonian 
\begin{eqnarray*}
H(k) = k^2 \Delta \sigma_+ \sigma_- + k \beta \sigma_+ +k \beta^\ast
\sigma_- +\omega_0,
\end{eqnarray*}
where $\sigma_-= | g \rangle \langle e |$, etc. Here $\Delta >0$ is
interpreted as a detuning parameter and $\beta $ as the amplitude of a
drive. Both the detuning and amplitude are assume to be large, which
corresponds to the limit $k\rightarrow \infty $.

The characteristic operator for the two-level system is then given by

\begin{eqnarray*}
\mathscr{T}_{k}\left( s\right) &=& \left[ 
\begin{array}{cc}
1 & 0 \\ 
0 & 1
\end{array}
\right] - \left[ 
\begin{array}{cc}
\sqrt{\gamma } & 0 \\ 
\sqrt{\kappa } & -\sqrt{\gamma }
\end{array}
\right] \left[ 
\begin{array}{cc}
s+\frac{1}{2}(\gamma +\kappa )+ik^{2}\Delta +i \omega_0  & - \frac{1}{2} \sqrt{\kappa \gamma }+ik\beta
\\ 
-  \frac{1}{2}\sqrt{\kappa \gamma }+ik\beta ^{\ast } & s+\frac{1}{2}\gamma
+i\omega _{0}
\end{array}
\right] ^{-1} \left[ 
\begin{array}{cc}
\sqrt{\gamma } & \sqrt{\kappa } \\ 
0 & -\sqrt{\gamma }
\end{array}
\right] ,  \nonumber
\end{eqnarray*}
which can be calculated explicitly as a $2\times 2$ matrix whose entries are
rational polynomials in $s$ of degree 2. What is of interest here is that
for large $k$ the characteristic operator takes the limit form 
\[
\lim_{k\rightarrow \infty } \mathscr{T}_{k}\left( s\right) =\left[ 
\begin{array}{cc}
1 & 0 \\ 
0 & \mathscr{T}_{g}(s)
\end{array}
\right] 
\]
where $\mathscr{T}_{g}\left( s\right) =\frac{s-\frac{1}{2}\gamma +i\omega
_{0}^{\prime }}{s+\frac{1}{2}\gamma +i\omega _{0}^{\prime }}$, where we have
the shifted frequency $\omega^\prime_0 = \omega _{0}-\frac{\left| \beta
\right| ^{2}}{\Delta }$. The limit model corresponds to the transfer
function of a linear system with a single degree of freedom having the
damping $\gamma $ and frequency $\omega^\prime_0 $.

What is happening in this limit is that the excited state plays an
increasingly negligible role in the model as its decay rate starts to
increase: the limit is a reduced model, however, with a shift of the
frequency.

\subsubsection{Qubit}

\label{subsubsec:qubit} As a next example, we consider a qubit driven by
three input fields, with 
\[
S=I_{3},L=\left[ 
\begin{array}{c}
\sqrt{\kappa _{1}}\sigma \\ 
\sqrt{\kappa _{2}}\sigma \\ 
\sqrt{\kappa _{3}}\sigma
\end{array}
\right] ,H=\Delta \sigma ^{\ast }\sigma -i\sqrt{\kappa _{1}}\left( \alpha
\sigma ^{\ast }-\alpha ^{\ast }\sigma \right) 
\]
where $\sigma ,\sigma ^{\ast }$ are the lowering and raising operators, $%
\Delta $ is a fixed detuning and $\alpha $ the amplitude of a drive field.
The characteristic operator now takes the form $\mathscr{T}\left( s\right) =\left[
\mathscr{T}_{jk}\left( s\right) \right] $ where we have the components 
\[
\mathscr{T}_{jk}\left( s\right) =\delta _{jk}I_{2}-\frac{\sqrt{\kappa _{j}\kappa _{k}}s%
}{s^{2}+(\frac{1}{2}\kappa +i\Delta )s+\kappa _{1}\left| \alpha \right| ^{2}}%
\sigma \sigma ^{\ast }, 
\]
for $j,k \in \{ 1,2,3 \}$ and where $\kappa =\kappa _{1}+\kappa _{2}+\kappa
_{3}$. In the special case where $\alpha =0$, there is a zero-pole
cancellation.

\section{Asymptotic Model reduction via Adiabatic Elimination}

We begin by considering the description of perturbations to open system
models in terms of their characteristic operators. We discuss regular
perturbations first for completeness: Suppose we have a model $\left(
S,L,H\right) $ which is a perturbation of solvable model $\left(
S,L,H_{0}\right) $ with 
\begin{eqnarray}
H=H_{0}+ \lambda V,
\end{eqnarray}
so that $K=-\frac{1}{2}L^{\ast }L-iH\equiv K_{0}-i\lambda V$. The resolvents 
$R(z)=\left( z-K\right) ^{-1}$ and $R_{0}(z)=\left( z-K_{0}\right) ^{-1}$
are then related by $R(z)=R_{0}(z)-i\lambda R(z)VR_{0}(z)$. For bounded
perturbation $V$ we have the Neumann series $R\left( z\right)
=\sum_{n=0}^{\infty }R_{0}\left( z\right) \left( -i\lambda VR_{0}\left(
z\right) \right) ^{n}$ so that the characteristic operators are related by 
\begin{eqnarray}
\mathscr{T}\left( z\right) =\mathscr{T}_{0}(z)-\sum_{n=1}^{\infty }
(-i\lambda )^n LR_{0}(z)\left( VR_{0}\left( z\right) \right) ^{n}L^{\ast }S.
\end{eqnarray}
This formula will be valid for suitably small constants $\lambda$. In
principle this formula may be useful for perturbative approaches to system
modelling.

\bigskip

Our main focus, however, will be singular perturbations corresponding to
adiabatic elimination.

\subsection{Fast and Slow Subspace Decomposition}

There exist a large body of results under the name of \textit{adiabatic
elimination} applicable to open quantum models. A universal mathematical
approach has been developed by Bouten, Silberfarb and van Handel \cite
{Bouten_Silberfarb,Bouten_Handel_Silberfarb}. We formulate their
presentation in a slightly different language. Essentially, the common
element in adiabatic elimination problems is that the system space can be
decomposed into a fast space, which is viewed as increasingly strongly
coupled to the bath, and a slow space. Specifically we assume a
decomposition of the system space as 
\begin{equation}
\mathfrak{h}=\mathfrak{h}_{\mathrm{slow}}\oplus \mathfrak{h}_{\mathrm{fast}}
\label{fs_decomp}
\end{equation}
A recent example of this is the approximate qubit regime for nonlinear
optical cavities \cite{Mabuchi_2012}. The coupling parameters are then taken
as $\left( S,L(k),H(k)\right) $ where $k$ is a strength parameter which we
eventually take to be large. For a given operator $X$ on $\mathfrak{h}$, we
write 
\begin{eqnarray}
X=\left[ 
\begin{array}{cc}
X_{\mathsf{ss}} & X_{\mathsf{sf}} \\ 
X_{\mathsf{fs}} & X_{\mathsf{ff}}
\end{array}
\right] .
\end{eqnarray}
More generally we use this notation when $X$ is an array of operators on $%
\mathfrak{h}$. The projections onto $\mathfrak{h}_{\mathrm{slow}}$ and $%
\mathfrak{h}_{\mathrm{fast}}$ are denoted by $P_{\mathsf{s}}\equiv \left[ 
\begin{array}{cc}
1 & 0 \\ 
0 & 0
\end{array}
\right] $ and $P_{\mathsf{f}}\equiv \left[ 
\begin{array}{cc}
0 & 0 \\ 
0 & 1
\end{array}
\right] $ respectively.

\subsection{Assumptions: Characteristic Operator Limit}

\label{assumptions}

\begin{enumerate}
\item  The coupling operator takes the form 
\begin{eqnarray}
L\left( k\right) =kL^{\left( 1\right) }+L^{\left( 0\right) }
\end{eqnarray}
where $L^{\left( 1\right) }P_{\mathsf{s}}=0$, that is, 
\begin{eqnarray}
L^{\left( 1\right) }\equiv \left[ 
\begin{array}{ll}
0 & L_{\mathsf{sf}}^{\left( 1\right) } \\ 
0 & L_{\mathsf{ff}}^{\left( 1\right) }
\end{array}
\right] ;
\label{eq:L(k)}
\end{eqnarray}

\item  The Hamiltonian takes the form $H\left( k\right) =H^{\left( 0\right)
}+kH^{\left( 1\right) }+k^{2}H^{\left( 2\right) }$ where $H^{\left( 1\right)
}P_{\mathsf{s}}=P_{\mathsf{s}}H^{\left( 1\right) }=0$ and $P_{\mathsf{s}%
}H^{\left( 2\right) }P_{\mathsf{s}}=0$, that is, 
\begin{eqnarray}
H\equiv \left[ 
\begin{array}{ll}
H_{\mathsf{ss}}^{\left( 0\right) } & H_{\mathsf{sf}}^{\left( 0\right) }+kH_{%
\mathsf{sf}}^{\left( 1\right) } \\ 
H_{\mathsf{fs}}^{\left( 0\right) }+kH_{\mathsf{fs}}^{\left( 1\right) } & H_{%
\mathsf{ff}}^{\left( 0\right) }+kH_{\mathsf{ff}}^{\left( 1\right) }+k^{2}H_{%
\mathsf{ff}}^{\left( 2\right) }
\end{array}
\right] ;
\end{eqnarray}

\item  In the expansion 
\begin{eqnarray}
K\left( k\right) =-\frac{1}{2}L(k)^{\ast }L(k)-iH(k)\equiv k^{2}A+kZ+R,
\label{eq:assumption_k}
\end{eqnarray}
we require that the operator 
\begin{eqnarray}
A_{\mathsf{ff}}=-\frac{1}{2}\sum_{\mathsf{a}=\mathsf{s},\mathsf{f}}L_{%
\mathsf{af}}^{\left( 1\right) \ast }L_{\mathsf{af}}^{\left( 1\right) }-iH_{%
\mathsf{ff}}^{\left( 2\right) }
\end{eqnarray}
be invertible on $\mathfrak{h}_{\mathsf{f}}$.
\end{enumerate}

\bigskip

Employing a repeated index summation convention over the index range $%
\left\{ \mathsf{s},\mathsf{f}\right\} $ from now on, we find that the
operator $R$ has components $R_{\mathsf{ab}}=-\frac{1}{2}L_{\mathsf{ca}%
}^{\left( 0\right) \ast }L_{\mathsf{cb}}^{\left( 0\right) }-iH_{\mathsf{ab}%
}^{\left( 0\right) }$ with respect to the slow-fast block decomposition.
Likewise 
\begin{eqnarray*}
A &\equiv& \left[ 
\begin{array}{ll}
0 & 0 \\ 
0 & A_{\mathsf{ff}}
\end{array}
\right] , \\
Z &\equiv& \left[ 
\begin{array}{ll}
0 & -\frac{1}{2}L_{\mathsf{cs}}^{\left( 0\right) \ast }L_{\mathsf{cf}%
}^{\left( 1\right) }-iH_{\mathsf{sf}}^{\left( 1\right) } \\ 
-\frac{1}{2}L_{\mathsf{cf}}^{\left( 1\right) \ast }L_{\mathsf{cs}}^{\left(
0\right) }-iH_{\mathsf{fs}}^{\left( 1\right) } & -\frac{1}{2}L_{\mathsf{cf}%
}^{\left( 0\right) \ast }L_{\mathsf{cf}}^{\left( 1\right) }-\frac{1}{2}L_{%
\mathsf{cf}}^{\left( 1\right) \ast }L_{\mathsf{cf}}^{\left( 0\right) }-iH_{%
\mathsf{ff}}^{\left( 1\right) }
\end{array}
\right] .  \label{K relations}
\end{eqnarray*}

In particular, we note the identities 
\begin{eqnarray}
R_{\mathsf{ss}}+R_{\mathsf{ss}}^{\ast } &=&-L_{\mathsf{cs}}^{\left( 0\right)
\ast }L_{\mathsf{cs}}^{\left( 0\right) }, \\
Z_{\mathsf{sf}}+Z_{\mathsf{fs}}^{\ast } &=&-L_{\mathsf{cs}}^{\left( 0\right)
\ast }L_{\mathsf{cf}}^{\left( 1\right) }, \\
A_{\mathsf{ff}}+A_{\mathsf{ff}}^{\ast } &=&-L_{\mathsf{cf}}^{\left( 1\right)
\ast }L_{\mathsf{cf}}^{\left( 1\right) }.  \label{K identities}
\end{eqnarray}

\subsection{The Characteristic Operator Limit}

In an adiabatic elimination problem, the coupling parameters $\left(
S,L\left( k\right) ,H\left( k\right) \right) $ lead to the associated
characteristic operator 
\begin{eqnarray}
\mathscr{T}_{k} ( s ) =S -L\left( k\right) \left[ s-K\left( k\right) \right]
^{-1}L\left( k\right) ^{\ast }S .
\end{eqnarray}


\begin{lemma}
\label{lem_limit}
Let $M\left( k\right) $ be a matrix parametrized by scalar $%
k$ of the form 
\[
M\left( k\right) =\left[ 
\begin{array}{cc}
M_{11} & kM_{12}+o\left( k\right) \\ 
kM_{21}+o\left( k\right) & k^{2}M_{22}+o\left( k\right)
\end{array}
\right] 
\]
with $M_{22}$ invertible. Then we have the limit 
\[
\lim_{k\rightarrow \infty }\left[ 
\begin{array}{cc}
1 & 0 \\ 
0 & k
\end{array}
\right] \left[ s+M\left( k\right) \right] ^{-1}\left[ 
\begin{array}{cc}
1 & 0 \\ 
0 & k
\end{array}
\right] =\left[ 
\begin{array}{ll}
\dfrac{1}{s+\widehat{M}_{11}} & -\dfrac{1}{s+\widehat{M}_{11}}M_{12}\dfrac{1}{M_{22}}
\\ 
-\dfrac{1}{M_{22}}M_{21}\dfrac{1}{s+\widehat{M}_{11}} & \quad \dfrac{1}{M_{22}}+%
\dfrac{1}{M_{22}}M_{21}\dfrac{1}{s+\widehat{M}_{11}}M_{12}\dfrac{1}{M_{22}}
\end{array}
\right] . 
\]
\end{lemma}

\noindent \textbf{Proof} see Appendix \ref{app:T_Lim}. $\square $

\begin{proposition}
\label{prop: limit} In the situation where the $L\left( k\right) $ and $%
H\left( k\right) $ are bounded operators for each $k$ fixed, the
characteristic operator has the strong limit 
\begin{eqnarray}
\widehat{\mathscr{T}}\left( s\right) =\lim_{k\rightarrow \infty }\mathscr{T}%
_{k}\left( s\right)  \label{lim}
\end{eqnarray}
for Re $s>0$, where we have

\[
\widehat{\mathscr{T}}_{\mathsf{ab}}\left( s\right) =\left\{ \delta _{\mathsf{ab}%
}+L_{\mathsf{af}}^{\left( 1\right) }\frac{1}{A_{\mathsf{ff}}}L_{\mathsf{cf}%
}^{\left( 1\right) \ast }-\left[ L_{\mathsf{as}}^{\left( 0\right) }-L_{%
\mathsf{af}}^{\left( 1\right) }\frac{1}{A_{\mathsf{ff}}}Z_{\mathsf{fs}}%
\right] \frac{1}{s-\widehat{K}_{\mathsf{ss}}}\left[ L_{\mathsf{cs}}^{\left(
0\right) \ast }-Z_{\mathsf{sf}}\frac{1}{A_{\mathsf{ff}}}L_{\mathsf{cf}%
}^{\left( 1\right) \ast }\right] \right\} S_{\mathsf{cb}}. 
\]

where 
\begin{equation}
\widehat{K}_{\mathsf{ss}}=R_{\mathsf{ss}}-Z_{\mathsf{sf}}\frac{1}{A_{\mathsf{ff}}%
}Z_{\mathsf{fs}}  .  \label{hat_K_ss}
\end{equation}
\end{proposition}

\noindent \textbf{Proof} This is a corollary to Lemma \ref{lem_limit}. In
this case we have the limit

\[
\lim_{k\rightarrow \infty }\left[ 
\begin{array}{cc}
1 & 0 \\ 
0 & k
\end{array}
\right] \left[ s-K\left( k\right) \right] ^{-1}\left[ 
\begin{array}{cc}
1 & 0 \\ 
0 & k
\end{array}
\right] =\left[ 
\begin{array}{ll}
\frac{1}{s-\widehat{K}_{\mathsf{ss}}}, & -\frac{1}{s-\widehat{K}_{\mathsf{ss}}}Z_{%
\mathsf{sf}}\frac{1}{A_{\mathsf{ff}}} \\ 
-\frac{1}{A_{\mathsf{ff}}}Z_{\mathsf{fs}}\frac{1}{s-\widehat{K}_{\mathsf{ss}}},
& \quad -\frac{1}{A_{\mathsf{ff}}}+\frac{1}{A_{\mathsf{ff}}}Z_{\mathsf{fs}}%
\frac{1}{s-\widehat{K}_{\mathsf{ss}}}Z_{\mathsf{sf}}\frac{1}{A_{\mathsf{ff}}}
\end{array}
\right] . 
\]
$\square $

\begin{proposition}
\label{prop:limit Top} The limit characteristic operator is given by 
\begin{eqnarray*}
\widehat{\mathscr{T}} =\left[ 
\begin{tabular}{l||l}
$-\frac{1}{2}\widehat{L}^{\ast }\widehat{L}-i\widehat{H}$ & $-\widehat{L}^{\ast }\widehat{S}$ \\ 
\hline\hline
$\widehat{L}$ & $\widehat{S}$%
\end{tabular}
\right] =\widehat{S}-\widehat{L}\left( s+\frac{1}{2}\widehat{L}^{\ast }\widehat{L}+i\widehat{H}%
\right) ^{-1}\widehat{L}^{\ast }\widehat{S}.
\end{eqnarray*}
where the parameters $(\widehat{S},\widehat{L},\widehat{H})$ are defined by 
\begin{eqnarray}
\widehat{S}=\left[ 
\begin{array}{cc}
\widehat{S}_{\mathsf{ss}} & \widehat{S}_{\mathsf{sf}} \\ 
\widehat{S}_{\mathsf{fs}} & \widehat{S}_{\mathsf{ff}}
\end{array}
\right] ,\;\widehat{L}=\left[ 
\begin{array}{cc}
\widehat{L}_{\mathsf{s}} & 0 \\ 
\widehat{L}_{\mathsf{f}} & 0
\end{array}
\right] ,\;\widehat{H}=\left[ 
\begin{array}{cc}
\widehat{H}_{\mathsf{ss}} & 0 \\ 
0 & 0
\end{array}
\right] ,
\end{eqnarray}
with 
\begin{eqnarray}
\widehat{S}_{\mathsf{ab}} &\triangleq &\left( \delta _{\mathsf{ac}}+L_{\mathsf{af%
}}^{\left( 1\right) }\frac{1}{A_{\mathsf{ff}}}L_{\mathsf{cf}}^{\left(
1\right) \ast }\right) S_{\mathsf{cb}}, \\
\widehat{L}_{\mathsf{a}} &\triangleq &L_{\mathsf{as}}^{\left( 0\right) }-L_{%
\mathsf{af}}^{\left( 1\right) }\frac{1}{A_{\mathsf{ff}}}Z_{\mathsf{fs}}, \\
\widehat{H}_{\mathsf{ss}} &\triangleq & H_{\mathsf{ss}}^{(0)}+\mathrm{Im}\left\{
Z_{\mathsf{sf}}\frac{1}{A_{\mathsf{ff}}}Z_{\mathsf{fs}}\right\} .
\end{eqnarray}
\end{proposition}

\noindent \textbf{Proof} See Appendix \ref{app:T_form}. $\square $

\subsection{Further assumptions}

We may impose additional constraints 
\begin{equation}
\widehat{L}_{\mathsf{f}}=\widehat{S}_{\mathsf{sf}}=\widehat{S}_{\mathsf{fs}}=0
\label{eqn_extra_conditions}
\end{equation}
to ensure that limit dynamics excludes the possibility of transitions that
terminate in any of the fast states. In this case $\widehat{S}_{\mathsf{ss}}$ is
unitary.

\begin{proposition}
If additionally (\ref{eqn_extra_conditions}) holds, then the slow and fast
subspaces are decoupled: 
\begin{eqnarray}
\widehat{\mathscr{T}}\left( s\right) =\left[ 
\begin{array}{cc}
\widehat{\mathscr{T}}_{\mathsf{ss}}\left( s\right) & 0 \\ 
0 & \widehat{S}_{\mathsf{ff}}
\end{array}
\right]
\end{eqnarray}
where 
\begin{eqnarray}
\widehat{\mathscr{T}}_{\mathsf{ss}}\left( s\right) =\left[ 
\begin{tabular}{l||l}
$-\frac{1}{2}\widehat{L}_{\mathsf{s}}^{\ast }\widehat{L}_{\mathsf{s}}-i\widehat{H}_{%
\mathsf{ss}}$ & $-\widehat{L}_{\mathsf{s}}^{\ast }\widehat{S}_{\mathsf{ss}}$ \\ 
\hline\hline
$\widehat{L}_{\mathsf{s}}$ & $\widehat{S}_{\mathsf{ss}}$%
\end{tabular}
\right] .
\end{eqnarray}
\end{proposition}

\noindent \textbf{Proof} This follows directly from 
\begin{eqnarray}
\widehat{\mathscr{T}}\left( s\right) =\left[ 
\begin{tabular}{l||l}
$\left[ 
\begin{array}{cc}
-\frac{1}{2}\widehat{L}_{\mathsf{s}}^{\ast }\widehat{L}_{\mathsf{s}}-i\widehat{H}_{%
\mathsf{ss}} & 0 \\ 
0 & 0
\end{array}
\right] $ & $-\left[ 
\begin{array}{cc}
\widehat{L}_{\mathsf{s}}^{\ast }\widehat{S}_{\mathsf{ss}} & 0 \\ 
0 & 0
\end{array}
\right] $ \\ \hline\hline
$\left[ 
\begin{array}{cc}
\widehat{L}_{\mathsf{s}} & 0 \\ 
0 & 0
\end{array}
\right] $ & $\left[ 
\begin{array}{cc}
\widehat{S}_{\mathsf{ss}} & 0 \\ 
0 & \widehat{S}_{\mathsf{ff}}
\end{array}
\right] $%
\end{tabular}
\right] .
\end{eqnarray}
$\square$

\subsection{Adiabatic Elimination for Quantum Stochastic Models}

The convergence of the characteristic operator is not sufficient to
guarantee the convergence of the corresponding unitary processes. In paper 
\cite{Bouten_Silberfarb} the extra condition (\ref{eqn_extra_conditions}) is
required.

\begin{theorem}[Bouten and Silberfarb 2008 \protect\cite{Bouten_Silberfarb}]

Suppose we are given a sequence of bounded operator parameters $\left(
S,L\left( k\right) ,H\left( k\right) \right) $ satisfying the assumptions in
equation (\ref{eqn_extra_conditions}). Then $U_{k}\left( t\right) P_{\mathsf{%
s}}$ converges strongly to $U\left( t\right) P_{\mathsf{s}}$, that is 
\begin{eqnarray}
\lim_{k\rightarrow \infty }\left\| U_{k}\left( t\right) \psi -U\left(
t\right) \psi \right\| =0
\end{eqnarray}
for all $\psi \in \mathfrak{h}\otimes \mathfrak{F}$ with $P_{\mathsf{f}%
}\otimes I\psi =0$.
\end{theorem}

The restriction to bounded coefficients was lifted in a subsequent
publication \cite{Bouten_Handel_Silberfarb}.

\subsection{Related Limits}

It is possible to consider more specific limits which may exist in
favourable circumstances. Foe instance, the all-pass form will lead to the
scaled $\Sigma $-function 
\[
\Sigma _{k}\left( s\right) =L\left( k\right) \frac{1}{s+iH\left( k\right) }%
L\left( k\right) ^{\ast } 
\]
which will converge provided $H_{\mathsf{ss}}^{\left( 2\right) }$ is
invertible on the slow space. In this case it happens the limit is
well-defined and given by 
\begin{eqnarray*}
\lim_{k\rightarrow \infty }\Sigma _{k}\left( s\right) &=&\left[ 
\begin{array}{ll}
0 & L_{\mathsf{sf}}^{\left( 1\right) } \\ 
0 & L_{\mathsf{ff}}^{\left( 1\right) }
\end{array}
\right] \left[ 
\begin{array}{ll}
\dfrac{1}{s+i\tilde{H}_{\mathsf{ss}}} & -\dfrac{1}{s+i\tilde{H}_{\mathsf{ss}}%
}H_{\mathsf{sf}}^{\left( 1\right) }\dfrac{1}{H_{\mathsf{ff}}^{\left(
2\right) }} \\ 
-\dfrac{1}{H_{\mathsf{ff}}^{\left( 2\right) }}H_{\mathsf{fs}}^{\left(
1\right) }\dfrac{1}{s+i\tilde{H}_{\mathsf{ss}}} & \quad -i\dfrac{1}{H_{%
\mathsf{ff}}^{\left( 2\right) }}+\dfrac{1}{H_{\mathsf{ff}}^{\left( 2\right) }%
}H_{\mathsf{fs}}^{\left( 1\right) }\dfrac{1}{s+i\tilde{H}_{\mathsf{ss}}}H_{%
\mathsf{sf}}^{\left( 1\right) }\dfrac{1}{H_{\mathsf{ff}}^{\left( 2\right) }}
\end{array}
\right] \left[ 
\begin{array}{ll}
0 & L_{\mathsf{sf}}^{\left( 1\right) } \\ 
0 & L_{\mathsf{ff}}^{\left( 1\right) }
\end{array}
\right] ^{\ast } \\
&=&\left[ 
\begin{array}{l}
L_{\mathsf{sf}}^{\left( 1\right) } \\ 
L_{\mathsf{ff}}^{\left( 1\right) }
\end{array}
\right] \left( -i\dfrac{1}{H_{\mathsf{ff}}^{\left( 2\right) }}+\dfrac{1}{H_{%
\mathsf{ff}}^{\left( 2\right) }}H_{\mathsf{fs}}^{\left( 1\right) }\dfrac{1}{%
s+i\tilde{H}_{\mathsf{ss}}}H_{\mathsf{sf}}^{\left( 1\right) }\dfrac{1}{H_{%
\mathsf{ff}}^{\left( 2\right) }}\right) \left[ 
\begin{array}{ll}
L_{\mathsf{sf}}^{\left( 1\right) \ast } & L_{\mathsf{ff}}^{\left( 1\right)
\ast }
\end{array}
\right] ,
\end{eqnarray*}
with $\tilde{H}_{\mathsf{ss}}=H_{\mathsf{ss}}^{\left( 1\right) }-H_{\mathsf{%
sf}}^{\left( 1\right) }\dfrac{1}{H_{\mathsf{ff}}^{\left( 2\right) }}H_{%
\mathsf{fs}}^{\left( 1\right) }$. We shall refer to this a the existence of
a limit in all pass. As we have seen, however, the general limit may exists
even when the Hamiltonian is zero.

More robust however, is the limit formulated in terms of the Stratonovich
form, where we have suitably-scaled coefficients $\mathbf{E}\left( k\right) $
and we use the Stratonovich form (\ref{eq:T_Strat}) along with Lemma \ref
{lem_limit}. 
We note the inverse relations 
\begin{eqnarray*}
E_{\ell \ell } &=&2i\frac{S-1}{S+1}, \\
E_{\ell 0} &=&\frac{2i}{S+1}L, \\
E_{00} &=&H+\frac{1}{4}L^{\ast }E_{\ell \ell }L.
\end{eqnarray*}

As $S$ is required to be $k$-independent, the same must be true for $E_{\ell
\ell }$. For convenience, we will fix the decompositions as $\mathfrak{h}=\mathfrak{h%
}_{\mathsf{s}}\oplus \mathfrak{h}_{\mathsf{f}}$ and assume that $E_{\ell \ell }$
is block diagonal:
\begin{eqnarray*}
E_{\ell \ell }\equiv \left[ 
\begin{array}{cc}
E_{\ell \ell }^{\left( \mathsf{s}\right) } & 0 \\ 
0 & E_{\ell \ell }^{\left( \mathsf{f}\right) }
\end{array}
\right] .
\end{eqnarray*}
Taking the form (\ref{eq:L(k)}) for $L\left( k\right) $, it follows that 
\begin{eqnarray*}
E_{\ell 0}\left( k\right) \equiv \left[ 
\begin{array}{cc}
0 & E_{\ell 0}^{\left( \mathsf{sf}\right) } \\ 
0 & E_{\ell 0}^{\left( \mathsf{ff}\right) }
\end{array}
\right] ,
\end{eqnarray*}
with $E_{\ell 0}^{\left( \mathsf{af}\right) }=i\left( 1+\frac{i}{2}E_{\ell
\ell }^{\left( \mathsf{a}\right) }\right) L_{\mathsf{af}}^{\left( 1\right) }$ (no summation!),
for $\mathsf{a}=\mathsf{s}$ or $\mathsf{f}$. It follows that in this case
\begin{eqnarray*}
E_{00}\left( k\right)  \equiv H\left( k\right) +\frac{k^{2}}{4}\left[ 
\begin{array}{cc}
0 & 0 \\ 
0 & E_{\ell 0}^{\left( \mathsf{sf}\right) \dag }E_{\ell \ell }^{\left( 
\mathsf{s}\right) }E_{\ell 0}^{\left( \mathsf{sf}\right) }+E_{\ell
0}^{\left( \mathsf{ff}\right) \dag }E_{\ell \ell }^{\left( \mathsf{f}\right)
}E_{\ell 0}^{\left( \mathsf{ff}\right) }
\end{array}
\right]  \equiv \left[ 
\begin{array}{cc}
E_{00}^{\left( \mathsf{ss}\right) } & kE_{00}^{\left( \mathsf{sf}\right) }
\\ 
kE_{00}^{\left( \mathsf{fs}\right) } & k^{2}E_{00}^{\left( \mathsf{ff}%
\right) }
\end{array}
\right] 
\end{eqnarray*}
which is again of the same form of the general matrix appearing in Lemma \ref{lem_limit}.
Provided that the self-adjoint term $E_{00}^{\left( \mathsf{ff}\right) }$ is
invertible on $\mathfrak{h}_{\mathsf{f}}$, the limit for the Stratonovich
expression exists and will agree with the previous limits. We omit the more general
situation where $E_{\ell \ell }$ is not block diagonal as it is more complicated
and not very enlightening.

\section{Hamiltonian Formulation of the Quantum Model}

In this section we describe how the unitary process $U(t)$ can alternatively
be viewed as Dirac picture unitaries relating a (singularly) perturbed
Hamiltonian dynamics to a free Hamiltonian dynamics.

\subsection{Dynamical Perturbations}

Let $V_{0}(t)$ and $V(t)$ be strongly continuous one-parameter groups, that
is $V_{0}(t+s)=V_{0}(t)V_{0}(s)$ and $V(t+s)=V(t)V(s)$, then we may view $V$
as a perturbed dynamics with respect to the free dynamics of $V_{0}$ by
transforming to the interaction picture via the wave operator 
\begin{eqnarray}
U(t)=V_{0}(t)^{\ast }V(t).
\end{eqnarray}
Physically $U(t)$ transforms to the Dirac picture \cite{Louisell}. It
inherits unitarity and strong continuity, but does not form a group. Instead
we have the so-called \emph{cocycle property} 
\begin{eqnarray}
U(t+s)=\Theta _{t}(U(s))U(t),
\end{eqnarray}
where $\Theta _{t}(x)=V_{0}(t)^{\ast }XV_{0}(t)$. By Stone's theorem, both $%
V_{0}$ and $V$ possess self-adjoint (Hamiltonian) infinitesimal generators $%
\widehat{H}_{0}$ and $\widehat{H}$ respectively: $i\dot{V}_{0}(t)=\widehat{H}%
_{0}V_{0}(t) $, and $i\dot{V}(t)=\widehat{H}V(t).$We say that $\widehat{H}$ is a
regular perturbation of $\widehat{H}_{0}$ if $\Upsilon =\widehat{H}-\widehat{H}_{0}$
defines an operator with dense domain. In this case, $U(t)$ will be strongly
differentiable and 
\begin{eqnarray}
i\dot{U}(t)=\Upsilon (t)U(t)
\end{eqnarray}
where the time-dependent Hamiltonian is $\Upsilon (t)=\Theta _{t}(\Upsilon )$%
. In situations where $\Upsilon $ is not densely defined, we will have a
singular perturbation and $U(t)$ will not generally be strongly
differentiable.

\subsection{Quantum Stochastic Evolutions}

The quantum input processes $b_{i}(t)$ may be view these processes a as
singular operators acting formally on the Hilbert space with the Fock space $%
\mathfrak{F}$ over $\mathbb{C}^{n}\otimes L^{2}(\mathbb{R})$. For $\Psi \in %
\mathfrak{F}$, we have a well-defined amplitude $\langle \tau
_{1},i_{1};\cdots ;\tau _{m},i_{m}|\Psi \rangle $ which is completely
symmetric under interchange of the $m$ pairs of labels $(\tau
_{1},i_{1}),\cdots ,(\tau _{m},i_{m})$, and this represent the amplitude to
have $m$ quanta with a particle of type $i_{1}$ at $\tau _{1}$, particle of
type $i_{2}$ at $\tau _{2}$, etc. We have the following resolution of
identity on $\mathfrak{F}$%
\begin{eqnarray}
\sum_{m=0}^{\infty }(\int d\tau _{1}\cdots d\tau
_{m})(\sum_{i_{1}=1}^{n}\cdots \sum_{i_{m}=1}^{n})\times |\tau
_{1},i_{1};\cdots ;\tau _{m},i_{m}\rangle \langle \tau _{1},i_{1};\cdots
;\tau _{m},i_{m}|=I.
\end{eqnarray}
The annihilator input process $b_{i}(t)$ is then defined almost everywhere
as 
\begin{eqnarray}
\langle \tau _{1},i_{1};\cdots ;\tau _{m},i_{m}|b_{i}(t)\Psi \rangle =\sqrt{%
m+1}\,\langle t,i;\tau _{1},i_{1};\cdots ;\tau _{m},i_{m}|\Psi \rangle .
\end{eqnarray}

The annihilation operators, together with their formal adjoints the creator
operators $b_{i}(t)^{\ast }$ satisfy the singular canonical commutation
relations $[b_{i}(t),b_{j}^\ast (s)]=\delta _{ij}\delta (t-s).$

\subsubsection{The Time Shift}

Let us introduce the following operator on the Fock space 
\begin{equation}
\widehat{H}_{0}=\sum_{j=1}^{n}\int_{-\infty }^{\infty }dt\,b^{\ast }(t)_{j}\,i%
\frac{\partial }{\partial t}\,b(t)_{j}  \label{K_0}
\end{equation}
which is the second quantization of the one-particle operator $i\frac{%
\partial }{\partial t}$. This is clearly a self-adjoint operator and the
unitary group $V_{0}(t)=e^{-it\widehat{H}_{0}}$ it generates is just the time
shift: 
\begin{eqnarray}
\langle \tau _{1},i_{1};\cdots ;\tau _{m},i_{m}|V_{0}(t)\Psi \rangle
=\langle \tau _{1}+t,i_{1};\cdots ;\tau _{m}+t,i_{m}|\Psi \rangle .
\end{eqnarray}
The free evolution $\Theta _{t}(\cdot )=V_{0}(t)^{\ast }(\cdot )V_{0}(t)$
will translate the input processes in time: $\Theta _{\tau
}(b_{i}(t))=b_{i}(t+\tau ),\;\Theta _{\tau }(b_{i}^{\ast }(t))=b_{i}^{\ast
}(t+\tau )$.

\subsection{Unitary QSDEs as Singular Perturbations}

The stochastic process $U(t)$ is strongly continuous, but due to the
presence of the noise fields $dB_{i}^{\ast },dB_{j}$ and $d\Lambda _{ij}$ is
not typically strongly differentiable. Here we see that the local
interaction $\Upsilon $ is a singular perturbation of the generator of the
time-shift $(\ref{K_0})$. We remark that nevertheless $U(t)$ is a $\Theta $%
-cocycle and that if we now define $V(t)$ by 
\begin{eqnarray}
V\left( t\right) =\left\{ 
\begin{array}{cc}
V_{0}\left( t\right) U\left( t\right) , & t\geq 0; \\ 
U\left( -t\right) ^{\ast }V_{0}\left( t\right) , & t<0.
\end{array}
\right.
\end{eqnarray}
then $V(t)$ is a strongly continuous unitary group and therefore admits an
infinitesimal generator $\widehat{H}$. Surprising as it may seem, the quantum
stochastic process $U(t)$ may be considered as the wave-operator for a
quantum dynamics with Hamiltonian $\widehat{H}$ with respect to the free
dynamics of the time shift generated by $\widehat{H}_{0}$. The relation 
\begin{eqnarray}
\widehat{H}=\widehat{H}_{0}+\Upsilon
\end{eqnarray}
however has only a formal meaning as the $\Upsilon $ is singular with
respect to $\widehat{H}_{0}$.

\subsection{Global Hamiltonian as Singular Perturbation of the Time Shift
Generator}

\label{sec:SingPert}

It has been a long standing problem to characterize the associated
Hamiltonian $\widehat{H}$\ for SLH models \cite{Acc90}. The major breakthrough
came in 1997 when A.N. Chebotarev solved this problem for the class of
quantum stochastic evolutions satisfying Hudson-Parthasarathy differential
equations with bounded commuting system coefficients \cite{Cheb97}. His
insight was based on scattering theory of a one-dimensional system with a
Dirac potential, say, with formal Hamiltonian 
\begin{eqnarray}
k=i\partial +E\delta
\end{eqnarray}
describing a one-dimensional particle propagating along the negative $x$%
-axis with a delta potential of strength $E$ at the origin. (In Chebotarev's
analysis the $\delta $-function is approximated by a sequence of regular
functions, and a strong resolvent limit is performed.) \ The mathematical
techniques used in this approach were subsequently generalized by Gregoratti 
\cite{Gregoratti} to relax the commutativity condition. More recently, the
analysis has been further extended to treat unbounded coefficients \cite{QG}.

Independently, several authors have been engaged in the program of
describing the Hamiltonian nature of quantum stochastic evolutions by
interpreting the time-dependent function $\Upsilon \left( t\right) $ as
being an expression involving quantum white noises satisfying a singular CCR 
\cite{Gough1,Gough99,AVL,VW}. This would naturally suggest that $\Upsilon $
should be interpreted as a sesquilinear expression in these noises at time $%
t=0$.

The generator of the free dynamics $k_{0}=i\partial $ is not semi-bounded
and the $\delta $-perturbation is viewed as a singular rank-one
perturbation. Here methods introduced by Albeverio and Kurasov \cite
{AlKur97,AlKur99a,AlKurBook00} may be employed to construct self-adjoint
extensions of such models, which we show in the next section for a wave on a
1-D wire.

\subsection{The Global Hamiltonian}

The form of the Hamiltonian $\widehat{H}$ is known to be \cite{Gregoratti} 
\begin{equation}
-i\widehat{H}\Psi =-i\tilde{H}_{0}\Psi -(\frac{1}{2}L_{i}^{\ast }L_{i}+iH)\Psi
-L_{i}^{\ast }S_{ij}b_{j}(0^{+})\Psi ,  \label{eq:Ham_def}
\end{equation}
on the domain of suitable functions satisfying the boundary condition 
\begin{equation}
b_{i}(0^{-})\Psi =L_{i}\Psi +S_{ij}\,b_{j}(0^{+})\Psi .  \label{eq:Ham_BC}
\end{equation}
here the suitable functions in question are those on the joint system and
Fock space that are in the domain of the free translation along the edges
(excluding the vertex at the origin) and in the domain of the one-sided
annihilators $b_{i}(0^{\pm })$. This agrees with the expression found in 
\cite{Cheb97} and \cite{Gregoratti}. The global Hamiltonian form is
essential for building up arbitrary quantum feedback networks \cite{GJ09}.

\subsection{Formal Linear System behind the $SLH$ Model}

We now specify to the case where the plant has finite dimensional Hilbert
space, say $\dim \mathfrak{h}=m<\infty $. In this case the operators $\left(
S,L,H\right) $ are naturally represented as complex-valued matrices with
dimensions 
\begin{eqnarray}
\mathsf{S}\in \mathbb{C}^{nm\times nm},\quad \mathsf{L}\in \mathbb{C}^{nm\times
m},\quad \mathsf{H}\in \mathbb{C}^{m\times m}.
\end{eqnarray}
That is, we have the matrix representations $\mathsf{S}_{ij},\mathsf{L}_{j},%
\mathsf{H}\in \mathbb{C}^{m\times m}$ for a fixed orthonormal basis of $h\cong 
\mathbb{C}^{m}$. In terms of the $\left( A,B,C,D\right) $ we then have 
\begin{eqnarray*}
A &=&\mathsf{K}=-\frac{1}{2}\sum_{j=1}^{n}\mathsf{L}_{j}^{\ast }\mathsf{L}%
_{j}-i\mathsf{H}\in \mathbb{C}^{m\times m}, \\
B &=&-\mathsf{L}^{\ast }\mathsf{S}=-[\sum_{j=1}^{n}\mathsf{L}_{j}^{\ast }%
\mathsf{S}_{j1},\cdots ,\sum_{j=1}^{n}\mathsf{L}_{j}^{\ast }\mathsf{S}%
_{jn}]\in \mathbb{C}^{m\times nm}, \\
C &=&\mathsf{L}=\left[ 
\begin{array}{c}
\mathsf{L}_{1} \\ 
\vdots \\ 
\mathsf{L}_{n}
\end{array}
\right] \in \mathbb{C}^{nm\times m}, \\
D &=&\mathsf{S}=\left[ 
\begin{array}{ccc}
\mathsf{S}_{11} & \cdots & \mathsf{S}_{1n} \\ 
\vdots & \ddots & \vdots \\ 
\mathsf{S}_{n1} & \cdots & \mathsf{S}_{nn}
\end{array}
\right] \in \mathbb{C}^{nm\times nm}.
\end{eqnarray*}
This is essentially the equivalent linear passive model considered in remark 
\ref{ref:passive_equiv}. Explicitly, the input-state-output equations behind
this will be 
\begin{eqnarray*}
\dot{x} &=&Ax+Bu \\
y &=&Cx+Du
\end{eqnarray*}
where $x$ is a $\mathbb{C}^{m}$-values state variable and $u$ and $y$ should be 
$\mathbb{C}^{nm}$-valued functions. Let $\Psi $ be a solution to the global
Hamiltonian problem (\ref{eq:Ham_def}) and satisfying the correct boundary
conditions (\ref{eq:Ham_BC}). This system may be rewritten as 
\begin{eqnarray}
\dot{\Psi}+i\tilde{H}_{0}\Psi &=&\mathsf{K}\Psi -\mathsf{L}^{\ast }\mathsf{S}%
u  \label{eq:Q1} \\
\quad \quad \quad \quad y &=&\mathsf{L}\Psi +\mathsf{S}u.  \label{eq:Q2}
\end{eqnarray}
where now the input and output functions are 
\begin{equation}
u_{j}=b_{j}(0^{+})\Psi ,\quad y_{j}=b_{j}(0^{-})\Psi  \label{eq:uy_BC}
\end{equation}
Absorbing the relatively unimportant free dynamics due to $\tilde{H}_{0}$,
we see that (\ref{eq:Q1},\ref{eq:Q2}) is linear system with ``input signal'' 
$u$ and ``output signal'' $y$.

The functions $u$ and $y$ are boundary terms related by (\ref{eq:uy_BC}) and
not to be interpreted literally as control functions which we can assign.

\section{Examples}

We now discuss some well-known examples from the perspective of control
theory.

\subsubsection{No scattering, and trivial damping}

Let us set $S=I$, $L^{(1)}=0$, and $L^{(0)}_{\mathsf{fs}}=0$. In this case
the only damping of significance is that of the slow component. Then we have 
$A_{\mathsf{ff}} = -i H^{(2)}_{\mathsf{ff}}$ and we require that $H^{(2)}_{%
\mathsf{ff}}$ is invertible on $\mathfrak{h}_{\mathsf{s}}$. It is easy to
see that the decoupling conditions now apply and we obtain the open dynamics
with $(\widehat{S}=I, \widehat{L}= L^{(0)}_{\mathsf{ss}}, \widehat{H})$ where the
reduced Hamiltonian is 
\[
\widehat{H}=H_{\mathsf{ss}}^{\left( 0\right) }-H_{\mathsf{sf} }^{(1)}\frac{1}{H_{%
\mathsf{ff}}^{(2)}}H_{\mathsf{fs}}^{\left( 1\right) } 
\]
Now $\widehat H$ is the shorted version (Schur complement) of $H(1)=\left[ 
\begin{array}{cc}
H_{\mathsf{ss}}^{\left( 0\right) } & H_{\mathsf{sf}}^{(1)} \\ 
H_{\mathsf{fs}}^{\left( 1\right) } & H_{\mathsf{ff}}^{(2)}
\end{array}
\right] $. Equivalently, $\widehat H$ is the the limit $k \uparrow \infty$ of
shorted version of $H(k)$.

The detuned two level atom model considered in subsection \ref
{subsubsec:detuned} is a special case.

\subsubsection{Qubit Limit}

Let us consider a cavity consisting of a single photon mode with annihilator 
$a$, so that $\left[ a,a^{\ast }\right] =I$. The number states $|n\rangle $, 
$\left( n=0,1,\cdots \right) $, span an infinite dimensional Hilbert space.
Mabuchi \cite{Mabuchi_2012} shows how a large Kerr non-linearity leads to a
reduced dynamics where we are restricted to the ground and first excited
state of the mode, and so have an effective qubit dynamics. We consider the $%
n=2$ input model with 
\begin{eqnarray*}
\left[ S(k)\right] _{jk} &=&\delta _{jk}I, \\
\left[ L\left( k\right) \right] _{j} &=&\sqrt{\kappa _{j}}e^{i\omega
t}a,\quad (j=1,2) \\
H\left( k\right) &=&k^{2}\chi _{0}a^{\ast 2}a^{2}+\Delta a^{\ast }a -i \sqrt{%
\kappa _{1}} \left( \alpha \left( t\right) a^{\ast }-\alpha ^{\ast }\left(
t\right) a\right) .
\end{eqnarray*}
In the model we are in a rotating frame with frequency $\omega $ and the
cavity is detuned from this frequency by a fixed amount $\Delta $. There is
a Kerr non-linearity of strength $\chi \left( k\right) =\chi _{0}k^{2}$
which will be the large parameter. We have two input fields with damping
rate $\kappa _{j} \, (j=1,2)$, and the first input introduces a coherent
driving field $\alpha \left( t\right) $.

We now have $A\equiv \chi _{0}a^{\ast 2}a^{2}=\chi _{0}N\left( N-1\right) $
where $N=a^{\ast }a$ is the number operator. The kernel space of $A$ is
therefore 
\[
\mathfrak{h}_{\mathrm{\mathsf{s}}}=\mathrm{span}\left\{ |0\rangle ,|1\rangle
\right\} . 
\]
For this situation we have $P_{\mathsf{s}}=|0\rangle \langle 0|+|1\rangle
\langle 1|$, and we find $L_{\mathsf{fs}}^{(0)}=0$ since $P_{\mathsf{f}}aP_{%
\mathsf{s}}\equiv 0$. The Bouten-Silberfarb conditions are then satisfied
and we have 
\[
H_{\mathsf{ss}}^{(0)}=P_{\mathsf{s}}\Delta a^{\ast }aP_{\mathsf{s}}\equiv
\Delta \sigma ^{\ast }\sigma 
\]
where $\sigma \triangleq P_{\mathsf{s}}aP_{\mathsf{s}}\equiv |0\rangle
\langle 1|$. We then have that 
\begin{eqnarray*}
\left[ \widehat{S}_{\mathsf{ss}}\right] _{jk} &=&\delta _{jk}\,I_{\mathsf{s}}, \\
\left[ \widehat{L}_{\mathsf{s}}\right] _{j} &=&\sqrt{\kappa _{j}}e^{i\omega
t}\sigma , \\
\widehat{H} &=&\Delta \sigma ^{\ast }\sigma -i\sqrt{\kappa _{1}}\left( \alpha
\left( t\right) \sigma ^{\ast }-\alpha ^{\ast }\left( t\right) \sigma
\right) .
\end{eqnarray*}
The system is then completely controllable through the control policy $\alpha $, and
observable through quadrature measurement (homodyning with $B_{\mathrm{out}%
,1}(t)-B_{\mathrm{out},1}(t)^{\ast }$, and $-iB_{\mathrm{out},1}(t)+iB_{%
\mathrm{out},1}(t)^{\ast }$) and by photon counting. The characteristic
operator is as computed in subsection \ref{subsubsec:qubit}. The limit
characteristic operator is then ($\kappa = \kappa_1 +\kappa_2$)
\[
\mathscr{T}_{\text{qubit}}\left( s\right) =\left[ 
\begin{array}{cc}
1 & 0 \\ 
0 & \frac{s^{2}-\frac{1}{2}\kappa s+i\Delta s+\kappa _{1}\left| \alpha
\right| ^{2}}{s^{2}+\frac{1}{2}\kappa s+i\Delta s+\kappa _{1}\left| \alpha
\right| ^{2}}
\end{array}
\right] . 
\]

\subsubsection{No scattering, but non-trivial damping}

We consider the case where $S=I$, $L_{\mathsf{fs}}^{(0)}=0$ and $L_{\mathsf{%
ff}}^{(1)}=0$, but $L_{\mathsf{sf}}^{(1)}\neq 0$. The decoupling conditions
are automatically satisfied, so all that is further required is that $A_{%
\mathsf{ff}}$, which is now given by 
\[
A_{\mathsf{ff}}\equiv -\frac{1}{2}L_{\mathsf{sf}}^{\left( 1\right) \ast }L_{%
\mathsf{sf}}^{\left( 1\right) }-iH_{\mathsf{ff}}^{\left( 2\right) }, 
\]
is invertible. If so the reduced $SLH$ takes the simplified form 
\begin{eqnarray*}
\widehat{S}_{\mathsf{ss}} &\equiv &I_{\mathsf{s}}+L_{\mathsf{sf}}^{\left(
1\right) }\frac{1}{A_{\mathsf{ff}}}L_{\mathsf{sf}}^{\left( 1\right) \ast },
\\
\widehat{L}_{\mathsf{s}} &\equiv &L_{\mathsf{ss}}^{\left( 0\right) }-L_{\mathsf{%
sf}}^{\left( 1\right) }\frac{1}{A_{\mathsf{ff}}}M_{\mathsf{fs}}, \\
\widehat{H} &\equiv &H_{\mathsf{ss}}^{(0)}+\mathrm{Im}\left\{ M_{\mathsf{sf}} 
\frac{1}{A_{\mathsf{ff}}}M_{\mathsf{fs}}\right\} ,
\end{eqnarray*}
where now 
\begin{eqnarray*}
M_{\mathsf{sf}} \equiv -\frac{1}{2}L_{\mathsf{ss}}^{\left( 0\right) \ast }L_{%
\mathsf{sf}}^{\left( 1\right) }-iH_{\mathsf{sf}}^{\left( 1\right) }, \quad
M_{\mathsf{fs}} \equiv -\frac{1}{2}L_{\mathsf{sf}}^{\left( 1\right) \ast }L_{%
\mathsf{ss}}^{\left( 0\right) }-iH_{\mathsf{fs}}^{\left( 1\right) }.
\end{eqnarray*}

\subsubsection{$\Lambda $-systems}

Consider a three level atom with ground states $|g_{1}\rangle $,$%
|g_{2}\rangle $ and an excited state $|e\rangle $ with Hilbert space $%
\mathfrak{h}_{\mathrm{level}}=\mathbb{C}^{3}$. The atom is contained in a
cavity with quantum mode $a$ with Hilbert space $\mathfrak{h}_{\mathrm{mode}%
} $ where $\left[ a,a^{\ast }\right] =1$ and $a$ annihilates a photon of the
cavity mode. The combined system and cavity has Hilbert space $\mathfrak{h}=%
\mathfrak{h}_{\mathrm{level}}\otimes \mathfrak{h}_{\mathrm{mode}}$, and
consider the following \cite{DuanKimble,Bouten_Handel_Silberfarb}, 
\begin{eqnarray*}
L\left( k\right) &=&k\sqrt{\gamma }I\otimes a, \\
H(k) &=&ik^{2}\mathsf{g}\left\{ |e\rangle \langle g_{1}|\otimes a-\mathrm{%
h.c.}\right\} +ik\left\{ |e\rangle \langle g_{2}|\otimes \alpha -\mathrm{h.c.%
}\right\} .
\end{eqnarray*}
Here the cavity is lossy and leaks photons with decay rate $\gamma $, we
also have a transition from $|e\rangle $ to $|g_{1}\rangle $ with the
emission of a photon into the cavity, and a scalar field $\alpha $ driving
the transition from $|e\rangle $ to $|g_{2}\rangle $. We see that 
\[
A\equiv -\frac{1}{2}\gamma I\otimes a^{\ast }a+\mathsf{g}\left\{ |e\rangle
\langle g_{1}|\otimes a-|g_{1}\rangle \langle e|\otimes a^{\ast }\right\} . 
\]
and that $A$ has a 2-dimensional kernel space spanned by the pair of states 
\[
|\Psi _{1}\rangle =|g_{1}\rangle \otimes |0\rangle ,\quad |\Psi _{2}\rangle
=|g_{2}\rangle \otimes |0\rangle . 
\]
The reduced subspace is then the span of $|\Psi _{1}\rangle $ and $|\Psi
_{2}\rangle $, and the resulting $SLH$ operators are 
\begin{eqnarray*}
\widehat{S} &=&|\Psi _{1}\rangle \langle \Psi _{1}|-|\Psi _{2}\rangle \langle
\Psi _{2}|\equiv I-2\sigma ^{\ast }\sigma , \\
\widehat{L} &=&-\frac{\gamma \alpha }{\mathsf{g}}|\Psi _{1}\rangle \langle \Psi
_{2}|\equiv -\frac{\gamma \alpha }{\mathsf{g}}\sigma , \\
\widehat{H} &=&0,
\end{eqnarray*}
where $\sigma =|\Psi _{1}\rangle \langle \Psi _{2}|$. Here the dynamics has
a vanishing Hamiltonian, but is partially observable through filtering as $%
\widehat{L}\neq 0$. The limit characteristic operator is then 
\[
\mathscr{T}_{\Lambda }\left( s\right) =\left[ 
\begin{array}{cc}
1 & 0 \\ 
0 & \frac{s-\frac{\gamma ^{2}}{2\mathsf{g}^{2}}\left| \alpha \right| ^{2}}{s+%
\frac{\gamma ^{2}}{2\mathsf{g}^{2}}\left| \alpha \right| ^{2}}
\end{array}
\right] . 
\]

\bigskip

Further examples of adiabatic elimination, particularly where the fast
degrees of freedom are oscillators, can be found in \cite{GNW,GN,Gough_RJMP}.

\section{Conclusions}

The characteristic operator is introduced here as a mathematical object
containing information about quantum input-output relations when processed
by a quantum mechanical system. The concept allows us to characterise
quantum systems, and many of the features associated with classical transfer
functions carry over. We have shown that it picks out the particular scaling
introduced by Bouten and Silberfarb for adiabatic elimination for quantum
open systems as being the one which leads to the convergence of
characteristic operators using Schur-Feshbach type resolvent expansions. It
is useful to note that strong coupling that restricts the degrees of freedom
adiabatically may also be interpreted as a projection onto a Zeno subspace,
though generally of an open systems character \cite{Gough_RJMP}.

We expect that the concept will play an important role in studying features
of quantum control systems such as model reduction, controllability and
observability.

\appendix

\section{Proof of Lemma \ref{lem_limit}}

\label{app:T_Lim} Again, by the Schur-Feshbach identity, we may write the
resolvent $\dfrac{1}{s+M\left( k\right) }$ as

\[
\left[ 
\begin{array}{cc}
s+M_{11}(k) & M_{12}(k) \\ 
M_{21}\left( k\right) & s+M_{22}\left( k\right)
\end{array}
\right] ^{-1}
=
\left[ 
\begin{array}{ll}
\Delta_{11}\left( s,k\right) & \Delta_{12}\left( s,k\right) \\ 
\Delta_{21}\left( s,k\right) & \Delta_{22}\left( s,k\right)
\end{array}
\right] 
\]
where, setting 
\begin{equation}
\widehat{M}_{11}\left( s,k\right)  \triangleq M_{11}(k)-M_{12} (k)\frac{1}{s+M_{22}\left( k\right) }M_{21}(k)
\end{equation}
we have 
\begin{eqnarray*}
\Delta_{11}\left( s,k\right) &=&\dfrac{1}{s+\widehat{M}_{11}(s,k)} \\
\Delta_{12}\left( s,k\right) &=&\widehat{M}_{11}\left( s,k\right) M_{12}\left(
k\right) \frac{1}{s+M_{22}(k)} \\
\Delta_{21}\left( s,k\right) &=&\frac{1}{s+M_{22}(k)}M_{21}\left( k\right) 
\widehat{M}_{11}\left( s,k\right) \\
\Delta_{22}\left( s,k\right) &=&\frac{1}{s+M_{22}(k)}+\frac{1}{s+M_{22}(k)}%
M_{21}\left( k\right) \dfrac{1}{s+\widehat{M}_{11}(s,k)}M_{12}\left( k\right) 
\frac{1}{s+M_{22}(k)}.
\end{eqnarray*}

Using the fact that $M_{12}\left( k\right) =kM_{12}+o(k),M_{21}\left(
k\right) =kM_{21}+o\left( k\right) $ and $M_{22}\left( k\right)
=k^2 M_{22}+o\left( k\right) $, we note
\begin{eqnarray*}
\widehat{M}_{11}\left( s,k\right) 
\equiv
M_{11}-k^{2}M_{12}\frac{1}{s+k^{2}M_{22}+o\left( k\right) }M_{21},  \label{K^hat (s,k) }
\end{eqnarray*}
and the following scaled limit 
\begin{eqnarray}
\widehat{M}_{11} \triangleq \lim_{k\rightarrow \infty }\widehat{M}_{11}\left( s,k\right)
=M_{11}-M_{12}\frac{1}{M_{22}}M_{21},
\end{eqnarray}
so that $\widehat{M}_{11}$ is a Schur complement of $\left[ 
\begin{array}{cc}
M_{11} & M_{12} \\ 
M_{21} & M_{22}
\end{array}
\right] $. Similarly it follows that

\[
\lim_{k\rightarrow \infty }\left[ 
\begin{array}{cc}
1 & 0 \\ 
0 & k
\end{array}
\right] \left[ s+M\left( k\right) \right] ^{-1}\left[ 
\begin{array}{cc}
1 & 0 \\ 
0 & k
\end{array}
\right] =\left[ 
\begin{array}{ll}
\dfrac{1}{s+\widehat{M}_{11}} & -\dfrac{1}{s+\widehat{M}_{11}}M_{12}\dfrac{1}{M_{22}}
\\ 
-\dfrac{1}{M_{22}}M_{21}\dfrac{1}{s+\widehat{M}_{11}} & \quad \dfrac{1}{M_{22}}+%
\dfrac{1}{M_{22}}M_{21}\dfrac{1}{s+\widehat{M}_{11}}M_{12}\dfrac{1}{M_{22}}
\end{array}
\right] . 
\]

\section{Proof of Proposition \ref{prop:limit Top}}

\label{app:T_form} Let us first note that we may define $\widehat{K}$ by $\widehat{K}%
=-\frac{1}{2}\widehat{L}^{\ast }\widehat{L}-i\widehat{H}$ in which case 
\begin{eqnarray}
\widehat{K}=\left[ 
\begin{array}{cc}
-\frac{1}{2}\widehat{L}_{\mathsf{a}}^{\ast }\widehat{L}_{\mathsf{a}}-i\widehat{H}_{%
\mathsf{ss}} & 0 \\ 
0 & 0
\end{array}
\right] .
\end{eqnarray}
We note that $-\frac{1}{2}\widehat{L}_{\mathsf{a}}^{\ast }\widehat{L}_{\mathsf{a}}-i%
\widehat{H}_{\mathsf{ss}}$ can be written as 
\begin{eqnarray*}
-\frac{1}{2}\left( L_{\mathsf{as}}^{\left( 0\right) \ast }-Z_{\mathsf{fs}%
}^{\ast }\frac{1}{A_{\mathsf{ff}}^{\ast }}L_{\mathsf{af}}^{\left( 1\right)
\ast }\right) \left( L_{\mathsf{as}}^{\left( 0\right) }-L_{\mathsf{af}%
}^{\left( 1\right) }\frac{1}{A_{\mathsf{ff}}}Z_{\mathsf{fs}}\right) -i\widehat{H}%
_{\mathsf{ss}} &=&R_{\mathsf{ss}}-\frac{1}{2}Z_{\mathsf{sf}}\frac{1}{A_{%
\mathsf{ff}}}Z_{\mathsf{fs}}+\frac{1}{2}Z_{\mathsf{fs}}^{\ast }\frac{1}{A_{%
\mathsf{ff}}^{\ast }}Z_{\mathsf{sf}}^{\ast } \\
&&-\frac{1}{2}\left( Z_{\mathsf{sf}}+Z_{\mathsf{fs}}^{\ast }\right) \frac{1}{%
A_{\mathsf{ff}}}Z_{\mathsf{fs}}-\frac{1}{2}Z_{\mathsf{fs}}^{\ast }\frac{1}{%
A_{\mathsf{ff}}^{\ast }}\left( Z_{\mathsf{sf}}+Z_{\mathsf{fs}}^{\ast }\right)
\\
&&+\frac{1}{2}Z_{\mathsf{fs}}^{\ast }\frac{1}{A_{\mathsf{ff}}^{\ast }}\left(
A_{\mathsf{ff}}+A_{\mathsf{ff}}^{\ast }\right) \frac{1}{A_{\mathsf{ff}}}Z_{%
\mathsf{fs}} \\
&=&R_{\mathsf{ss}}-Z_{\mathsf{sf}}\frac{1}{A_{\mathsf{ff}}}Z_{\mathsf{fs}}
\end{eqnarray*}
where we use $\left( \ref{K identities}\right) $.

Therefore, with $\widehat{K}_{\mathsf{ss}}$ is as defined in $\left( \ref
{hat_K_ss}\right) $, we have 
\begin{eqnarray}
\widehat{K}=\left[ 
\begin{array}{cc}
\widehat{K}_{\mathsf{ss}} & 0 \\ 
0 & 0
\end{array}
\right] .
\end{eqnarray}
Moreover, we see that $\widehat{S}$ is unitary. To see this, set $\widehat{T}=\widehat{S}%
S^{-1}$ then 
\begin{eqnarray*}
\widehat{T}_{\mathsf{ca}}^{\ast }\widehat{T}_{\mathsf{cb}}=[\delta _{\mathsf{ca}}+L_{%
\mathsf{af}}^{\left( 1\right) }\frac{1}{A_{\mathsf{ff}}^{\ast }}L_{\mathsf{cf%
}}^{\left( 1\right) \ast }][\delta _{\mathsf{cb}}+L_{\mathsf{cf}}^{\left(
1\right) }\frac{1}{A_{\mathsf{ff}}}L_{\mathsf{bf}}^{\left( 1\right) \ast }]
=\delta _{\mathsf{ab}}+L_{\mathsf{af}}^{\left( 1\right) }\frac{1}{A_{\mathsf{%
ff}}^{\ast }}\left\{ A_{\mathsf{ff}}+A_{\mathsf{ff}}^{\ast }+L_{\mathsf{cf}%
}^{\left( 1\right) \ast }L_{\mathsf{cf}}^{\left( 1\right) }\right\} \frac{1}{%
A_{\mathsf{ff}}}L_{\mathsf{bf}}^{\left( 1\right) \ast }
\end{eqnarray*}
however the expression in braces vanishes identically leaving $\widehat{T}^{\ast
}\widehat{T}=I$. The proof of the co-isometric property of $\widehat{T}\widehat{T}^{\ast
}=I$ is similar.

We note that 
\begin{eqnarray}
\widehat{S}\equiv \lim_{|s|\rightarrow \infty }\widehat{\mathscr{T}}\left( s\right) .
\end{eqnarray}

It remains to show that the limit characteristic function $\widehat{\mathscr{T}}$
has the stated form. Substituting in form $\left( \ref{lim}\right) $, we have

\begin{eqnarray*}
\widehat{\mathscr{T}}_{\mathsf{ab}}\left( s\right) -\left[ \widehat{S}_{\mathsf{ab}}-%
\widehat{L}_{\mathsf{a}}\left( s-\widehat{K}_{\mathsf{ss}}\right) ^{-1}\widehat{L}_{%
\mathsf{c}}\widehat{S}_{\mathsf{cb}}\right] =\widehat{L}_{\mathsf{a}}\frac{1}{s-\widehat{%
K}_{\mathsf{ss}}}\left\{ -(L_{\mathsf{ds}}^{\left( 0\right) }-Z_{\mathsf{sf}}%
\frac{1}{A_{\mathsf{ff}}}L_{\mathsf{df}}^{\left( 1\right) \ast })+\widehat{L}_{%
\mathsf{c}}^{\ast }(\delta _{\mathsf{cd}}+L_{\mathsf{cf}}^{\left( 1\right) }%
\frac{1}{A_{\mathsf{ff}}}L_{\mathsf{df}}^{\left( 1\right) \ast })\right\} S_{%
\mathsf{db}}
\end{eqnarray*}
and the term in braces equals 
\begin{eqnarray}
\left[ Z_{\mathsf{sf}}\frac{1}{A_{\mathsf{ff}}}-Z_{\mathsf{fs}}^{\ast }\frac{%
1}{A_{\mathsf{ff}}}+L_{\mathsf{cs}}^{\left( 0\right) \ast }L_{\mathsf{cf}%
}^{\left( 1\right) }\frac{1}{A_{\mathsf{ff}}}-L_{\mathsf{fs}}^{\left(
1\right) \ast }\frac{1}{A_{\mathsf{ff}}}L_{\mathsf{cf}}^{\left( 1\right)
\ast }L_{\mathsf{cf}}^{\left( 1\right) }\frac{1}{A_{\mathsf{ff}}}\right] L_{%
\mathsf{df}}^{\left( 1\right) \ast }
\end{eqnarray}
and using $\left( \ref{K identities}\right) $ again we see that the term in
square brackets is 
\begin{eqnarray}
Z_{\mathsf{sf}}\frac{1}{A_{\mathsf{ff}}}-Z_{\mathsf{fs}}^{\ast }\frac{1}{A_{%
\mathsf{ff}}}-\left( Z_{\mathsf{sf}}+Z_{\mathsf{fs}}^{\ast }\right) \frac{1}{%
A_{\mathsf{ff}}}-Z_{\mathsf{fs}}^{\ast }\frac{1}{A_{\mathsf{ff}}}\left( A_{%
\mathsf{ff}}+A_{\mathsf{ff}}\right) \frac{1}{A_{\mathsf{ff}}}
\end{eqnarray}
which vanishes identically.

\bigskip

We note that we have the alternative form 
\begin{eqnarray}
\widehat{H}_{\mathsf{ss}}=H_{\mathsf{ss}}^{\left( 0\right) }-Z_{\mathsf{fs}%
}^{\ast }\frac{1}{A_{\mathsf{ff}}}H_{\mathsf{fs}}^{\left( 1\right) }-H_{%
\mathsf{sf}}^{\left( 1\right) }\frac{1}{A_{\mathsf{ff}}}Z_{\mathsf{fs}}+Z_{%
\mathsf{fs}}^{\ast }\frac{1}{A_{\mathsf{ff}}}H_{\mathsf{ff}}^{\left(
2\right) }\frac{1}{A_{\mathsf{ff}}}Z_{\mathsf{fs}}.
\end{eqnarray}

\section*{Acknowledgement}

This research was supported in part by the National Science Foundation under
Grant No. NSF PHY11-25915, and the author is grateful for the support of the
Kavli Institute for Theoretical Physics, U.C. Santa Barbara, where this work
begun during the Control of Complex Quantum Systems programme in January
2013. It was also supported by EPSRC grant EP/L006111/1, and the author is
particularly grateful to Dr Hendra Nurdin for several valuable technical
comments while writing this paper. Finally he as the pleasant duty to thank
the Isaac Newton Institute for Mathematical Sciences, Cambridge, for support
and hospitality during the programme \textit{Quantum Control Engineering}
August 2014 where work on this paper was completed. He acknowledges several
fruitful discussions with Luc Bouten, Jake Taylor, Matthew James and Gerard
Milburn.


\end{document}